\documentclass[10pt,twocolumn,floatfix,prb,superscriptaddress]{revtex4-2}
\usepackage{amsmath,amssymb,amsthm,mathrsfs,amsfonts,dsfont,amstext} 
\usepackage{textcomp,pbox}
\usepackage[export]{adjustbox}
\usepackage{soul} 
\usepackage{bm,xspace}
\usepackage{dcolumn,booktabs,url}
\usepackage[scaled]{helvet}
\usepackage{sansmath,gensymb}
\usepackage{tikz,graphicx,transparent,color}
\usepackage{multirow}
\usepackage[separate-uncertainty = true]{siunitx}
\usepackage{comment}
\usepackage{physics}
\usepackage{pdfpages}
\usepackage{array}
\usepackage{makecell}
\usepackage{tabularx}
\usepackage{dsfont}

\newcolumntype{L}{>{\raggedright\arraybackslash}X}

\makeatletter
\AtBeginDocument{\let\LS@rot\@undefined}
\makeatother

\newcolumntype{C}[1]{>{\middleing\let\newline\\\arraybackslash\hspace{0pt}}m{#1}}

\usepackage{natbib} 

\usepackage[colorlinks=true]{hyperref}
\usepackage{graphicx}

\hypersetup{
     colorlinks   = true,
     citecolor    = red,
     linkcolor    = blue,
     urlcolor     = red     
}

\graphicspath{{./figs/}}

\newcommand{\SM}{Supplementary\xspace}
\newcommand{\figref}[1]{\figurename{~\ref{#1}}}

\newcommand\numberthis{\addtocounter{equation}{1}\tag{\theequation}} 

\makeatletter
\def\maketitle{
\@author@finish
\title@column\titleblock@produce
\suppressfloats[t]}
\makeatother

\usepackage[resetlabels]{multibib}

\newcites{S}{References}

\usepackage{lineno}

\begin{document}

\newcommand{\TitleName}{Violation of Bell inequality by photon scattering on a two-level emitter}

\title{\TitleName}

\newcommand{\AffCPH}{Center for Hybrid Quantum Networks (Hy-Q), The Niels Bohr Institute, University~of~Copenhagen,  DK-2100  Copenhagen~{\O}, Denmark}
\newcommand{\AffBasel}{Department of Physics, University of Basel, Klingelbergstra\ss e 82, CH-4056 Basel, Switzerland}
\newcommand{\AffBochum}{Lehrstuhl f\"ur Angewandte Fest\"orperphysik, Ruhr-Universit\"at Bochum, Universit\"atsstra\ss e 150, 44801 Bochum, Germany}

\author{Shikai Liu}
\affiliation{\AffCPH{}}
\author{Oliver August Dall'Alba  Sandberg}
\affiliation{\AffCPH{}}
\author{Ming Lai Chan}
\affiliation{\AffCPH{}}
\author{Bj\"{o}rn Schrinski}
\affiliation{\AffCPH{}}
\author{Yiouli Anyfantaki}
\affiliation{\AffCPH{}}
\author{Rasmus Bruhn Nielsen}
\affiliation{\AffCPH{}}
\author{Robert Garbecht Larsen}
\affiliation{\AffCPH{}}
\author{Andrei Skalkin}
\affiliation{\AffCPH{}}
\author{Ying Wang}
\affiliation{\AffCPH{}}
\author{Leonardo Midolo}
\affiliation{\AffCPH{}}
\author{Sven Scholz}
\affiliation{\AffBochum{}}
\author{Andreas Dirk Wieck}
\affiliation{\AffBochum{}}
\author{Arne Ludwig}
\affiliation{\AffBochum{}}

\author{Anders Søndberg S\o{}rensen}
\affiliation{\AffCPH{}}
\author{Alexey Tiranov$^{\dag}$}
\thanks{Present address: Chimie ParisTech, Université PSL, CNRS, Institut de Recherche de Chimie Paris, 75005 Paris, France}
\affiliation{\AffCPH{}}

\author{Peter Lodahl}
\thanks{Email to: alexey.tiranov@chimieparistech.psl.eu; lodahl@nbi.ku.dk}
\affiliation{\AffCPH{}}

\date{\today}

\begin{abstract}

Entanglement, the non-local correlations present in multipartite quantum systems, is a curious feature of quantum mechanics and the fuel of quantum technology. It is therefore a major priority to develop energy-conserving and simple methods for generating high-fidelity entangled states. In the case of light, entanglement can be realized by interactions with matter, although the required nonlinear interaction is typically weak, thereby limiting its applicability. Here, we show how a single two-level emitter deterministically coupled to light in a nanophotonic waveguide is used to realize genuine photonic quantum entanglement for excitation at the single photon level. By virtue of the efficient optical coupling, two-photon interactions are strongly mediated by the emitter realizing a giant nonlinearity that leads to entanglement. We experimentally generate and verify energy-time entanglement by violating a Bell inequality (Clauder-Horne-Shimony-Holt Bell parameter of $S=2.67(16)>2$) in an interferometric measurement of the two-photon scattering response. As an attractive feature of this approach, the two-level emitter acts as a passive scatterer initially prepared in the ground state, i.e., no advanced spin control is required. This experiment is a fundamental advancement that may pave a new route for ultra-low energy-consuming synthesis of photonic entangled states for quantum simulators or metrology.

\end{abstract}

\maketitle 

\begin{figure*}[hbtp!]
	\includegraphics[width=0.95\linewidth, trim=0.cm 0.05cm 0.cm 0.0cm,clip]{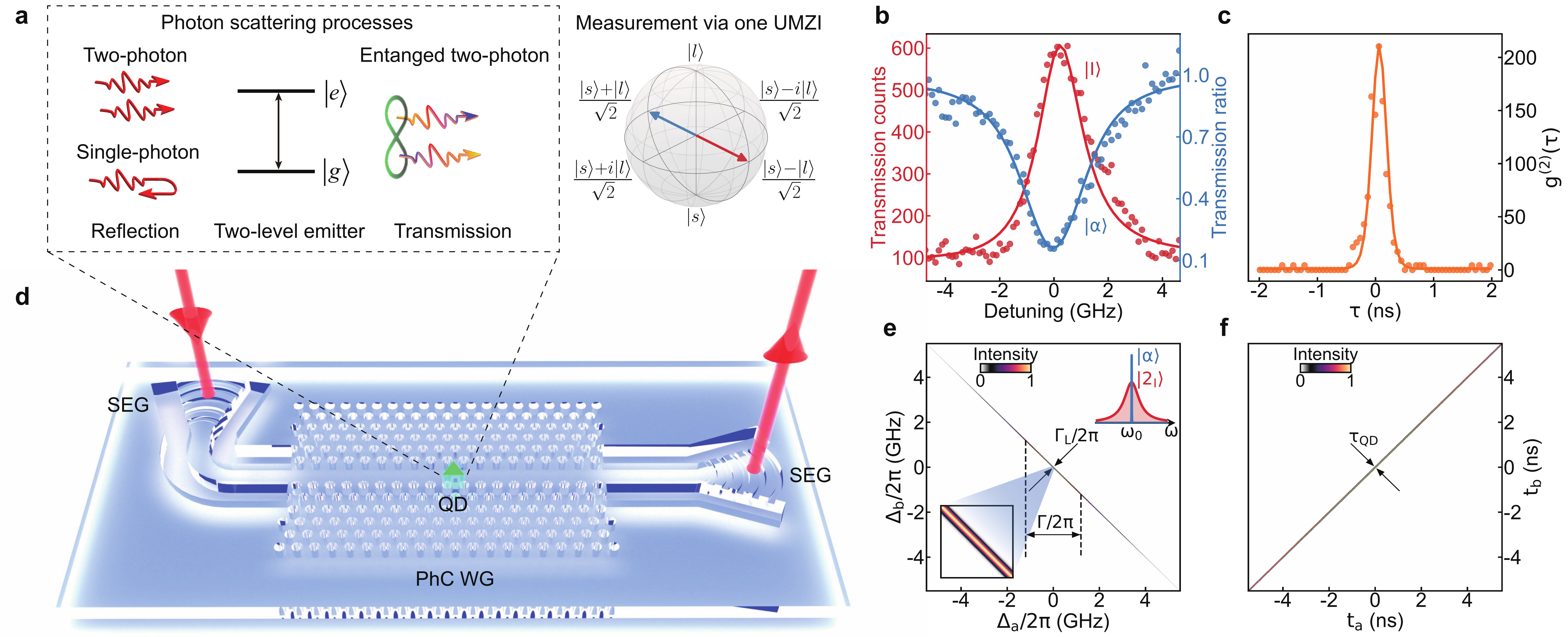}
	\caption{ 
	(color online) \textbf{Two-photon energy-time entanglement induced by coherent interaction of two photons with a quantum dot (QD) integrated into a photonic crystal waveguide (PhC WG).}
    (a) Operational principle of the photon scattering and entanglement processes. A single-photon wave packet is predominantly reflected by elastic scattering on a two-level emitter, while the two-photon wave packet can be inelastically scattered in the forward direction, thereby generating energy-time entanglement. The entanglement is probed using two unbalanced Mach–Zehnder interferometers (UMZIs), see \figref{fig:2}. Each UMZI is used to realize time projections onto the superposition state $\frac{\ket{s}+ e^{i\phi}\ket{l}}{\sqrt2}$, as illustrated on a Bloch sphere, where $\ket{s}$ ($\ket{l}$) corresponds to a photon taking the short (long) path, and $\phi$ is the phase setting of the UMZI. $\phi=0$ (blue vector) and $\phi=\pi$ (red vector) refer to the two settings for data in (b).  
    (b) Transmission intensity measurements through the PhC WG and one UMZI versus QD detuning. The blue curve indicates the transmission dip by resonant scattering of a weak coherent state $\ket{\alpha}$. Suppression of the elastically scattered laser photons by a destructive interference phase ($\phi=\pi$) reveals inelastically scattered photons $\ket{I}$ (red). 
    (c) Measured normalized second-order correlation function $g^{(2)}(\tau)$ of the light transmitted through the PhC WG on resonance with the QD, reaching values above 200, a signature of strong bunching induced by the scattering.
    (d) Schematic of the QD-embedded PhC WG structure with two mode adaptors, including shallow-etched gratings (SEGs) and nanobeam access waveguides to the photonic crystal section. 
    (e) Calculated normalized joint spectral intensity for a laser linewidth of $\Gamma_L/2\pi=100$ kHz, a Purcell enhanced QD linewidth of $\Gamma/2\pi=2.3$ GHz, and assuming ideal coupling of the QD to the PhC WG. $\Delta_{a(b)}$ is the frequency difference between the output photon ($a \text{ or } b$) and the input laser: $\Delta_{a(b)}=\omega_{a(b)}-\omega_p$ . The width of the biphoton spectrum is determined by the laser linewidth $\Gamma_L$, while each photon is broadened by the QD linewidth $\Gamma$. Upper-right insert: the corresponding spectra of the input coherent state $\ket{\alpha}$ (blue) and the output biphoton state $\ket{2_I}$ (red), respectively. Bottom-left insert: enlarged joint spectral intensity spanning a range of 1 MHz.
    (f) The corresponding normalized joint temporal intensity. The biphoton correlation time is determined by the QD lifetime $\tau_{QD}$, cf. \SM~Note~\ref{lifetime} for characterization. 
}
	\label{fig:1}
\end{figure*}

A quantum pulse of light interacting with a two-level emitter, see \figref{fig:1}(a,d), constitutes a new experimental paradigm in quantum optics \cite{Shen2007,Kiilerich2019}. Despite its conceptual simplicity, significant quantum complexity can be encoded in the system since a quantum pulse represents an infinitely large (continuous) Hilbert space. From an experimental point of view, this is an attractive setting since a two-level emitter can implement a highly nonlinear operation on the incoming pulse without the need for demanding and error-susceptible emitter preparation schemes. To enhance this photon-photon nonlinearity, the main experimental challenge is to promote the radiative coupling of the emitter such that it dominates deteriorating decoherence processes -- such advancements have been made in the past decades using semiconductor quantum dots (QDs) in photonic crystal waveguides (PhC WGs) and cavities~\cite{Lodahl2022};
see illustration of a PhC WG device in \figref{fig:1}(d). 

It is an exciting ongoing research endeavor to exploit the properties and applicability of the quantum nonlinear response of a two-level emitter. The essential nonlinear operation has been explored using various quantum emitters such as QDs \cite{Javadi2015a},
color centers in diamond~\cite{Sipahigil2016},  atoms~\cite{Hamsen2017}, and single molecules~\cite{Pscherer2021}. 
Further experimental advancements include quadrature squeezing of light~\cite{Schulte2015,Hinney2021}, two-photon correlation dynamics and photonic bound states~\cite{Jeannic2022,Tomm2023}. Theoretical proposals include photon sorters for deterministic Bell state analyzers~\cite{Witthaut2012}, quantum logic gates~\cite{Konyk2019,Krastanov2022}, and single-photon transistors~\cite{Chang2007}. Furthermore, many-body waveguide quantum electrodynamics may be pushed to new realms of strongly correlated light and matter~\cite{Fayard2021,Bello2022}.
While it was theoretically predicted that the two-level nonlinear response can induce photon-photon correlations~\cite{Shen2007}, whether this nonlinearity can lead to photon entanglement has never been truly verified experimentally.

Here, we demonstrate that a two-level quantum emitter coherently coupled to a PhC WG can induce strong energy-time entanglement between two scattered photons, see \figref{fig:1}(a), sufficient for violating a Bell inequality and thus local realism under the fair sampling assumption.
The experiment couples a continuous wave (incoming light) and a discrete quantum system (emitter), offering a route to non-Gaussian photonic operations -- a type of operation vigorously searched for in continuous-variable quantum computing architectures~\cite{Bourassa2021}.
Previous research on entanglement generation with quantum emitters exploited the strong excitation (Mollow) regime in bulk samples~\cite{Peiris2017a} or the QD biexciton radiative cascade~\cite{Muller2008,Jayakumar2013,Hohn2023}. To the best of our knowledge, no such studies have yet answered the above open question and proved experimentally that passive scattering of weak fields from a two-level QD in a PhC WG can induce genuine entanglement. This work opens a conceptually new and advantageous route to energy-time entanglement generation that may be an attractive alternative to four-wave mixing sources~\cite{Glorieux2010}  since it operates at the ultra-low energy consumption level of single photons and no complex and decoherence-sensitive pumping schemes are required.

\begin{figure*}[hbtp!]
	\includegraphics[width=0.95\linewidth, trim=0.8cm 0.4cm 0.4cm 0.4cm,clip]{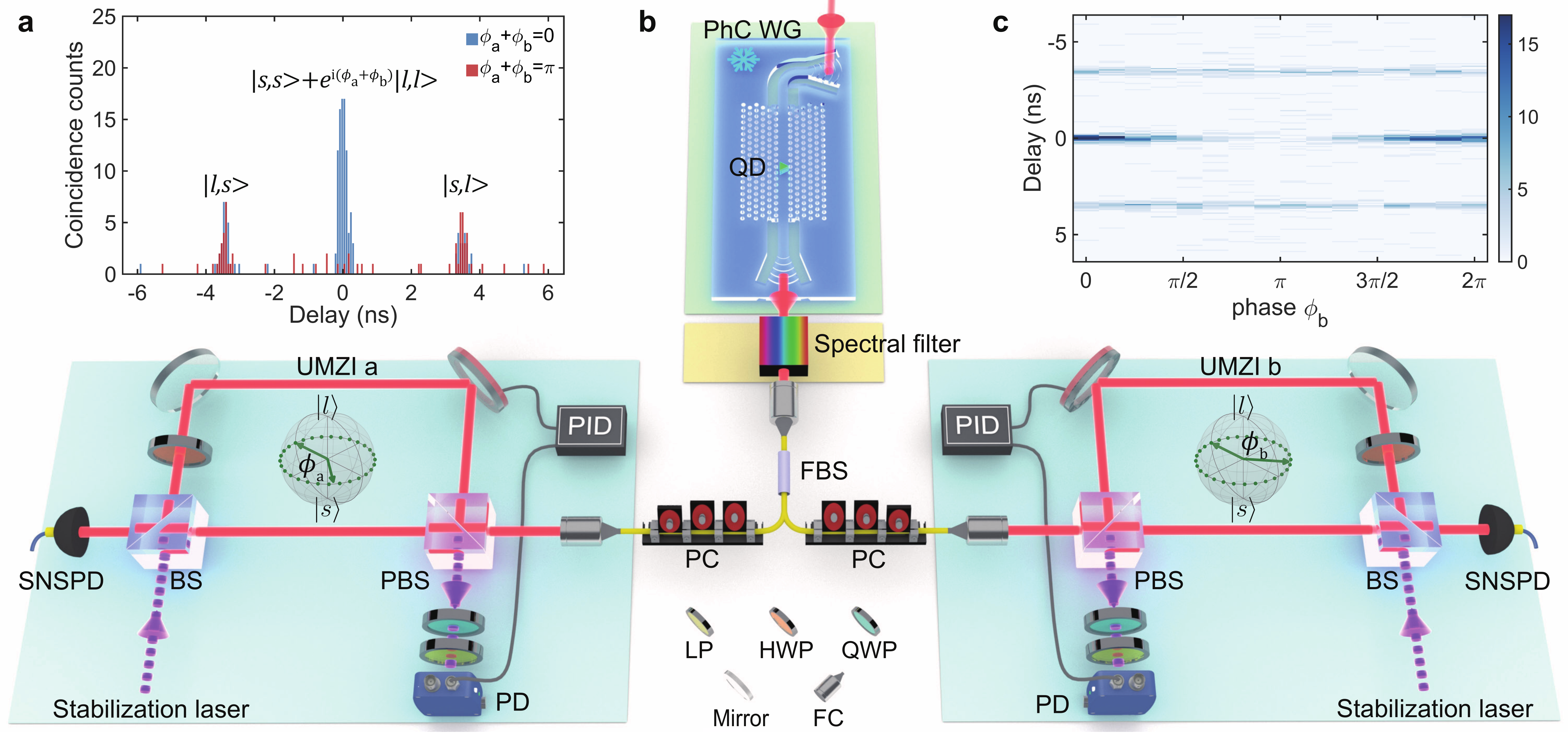}
	\caption{ 
    	(color online) \textbf{Experimental setup and characterization of two-photon energy-time entanglement}. 
    (a) Time correlation histograms of coincidence counts for constructive (blue, $\phi_a+\phi_b=0$) and destructive (red, $\phi_a+\phi_b=\pi$) interference between two photons traversing the short ($s$) and long ($l$) arms of the UMZIs, respectively.
    (b) Experimental setup including the PhC WG chip (light green area), spectral filter (light orange area), and the Franson interferometer (light blue areas). The two Bloch spheres illustrate the two independently controlled phases $\phi_a$ and $\phi_b$. LP: linear polarizer; HWP: half-wave plate; QWP: quarter-wave plate;  SNSPD: superconducting nanowire single-photon detector; FC: fiber collimator; BS: beam splitter; PBS: polarizing beam splitter; PC, polarization controller; FBS, fiber beam splitter; PD: photodiode. In order to control the phase difference between the two interferometer arms, we actively stabilize the UMZIs with a PID module locked by the same laser that excites the QD.
    (c) 2D correlation histogram of coincidence counts versus phase and time delay. 
}
	\label{fig:2}
\end{figure*}

We consider a single two-level emitter deterministically coupled to a single propagating spatial mode in a PhC WG, cf. \figref{fig:1}(d). A weak coherent input field is launched into the PhC WG and interacts with the emitter of coupling efficiency $\beta= \frac{\gamma}{\Gamma}$ governed by the ratio between the radiative decay rate into the waveguide mode $\gamma$ and the QD total decay rate $\Gamma$~\cite{Lodahl2015}. For $\beta=1$ with no decoherence processes, the single-photon component is elastically reflected via interaction with the emitter, while the two-photon component can be inelastically scattered into the forward (transmission) mode, leading to energy exchange and photon bunching in time~\cite{Shen2007,Chang2007}, as shown in \figref{fig:1}(a). The latter process is analogous to degenerate four-wave mixing with a Kerr nonlinearity~\cite{Glorieux2010}. Experimentally, the two-photon scattering process is studied in a Franson interferometer~\cite{Franson1989} with time-resolved photon correlation measurements, see \figref{fig:2}. In this way, a Clauder-Horn-Shimony-Holt (CHSH) Bell inequality entanglement criterion can be tested where a Bell parameter of $S=2$ constitutes the locality bound~\cite{Clauser1969}. Various experimental imperfections influence $S$, including the finite photon-emitter coupling efficiency ($\beta$-factor), pure dephasing rate ($\gamma_d$ relative to the emitter linewidth $\Gamma$), and the strength of the incoming light (mean photon number within the emitter lifetime $n$). These imperfections result in a single-photon component (elastic scattering) that is not fully reflected thereby reducing $S$. We found that $S$ is sensitive to $\gamma_d$, $\Gamma$ and $n$ to first-order but is remarkably robust to coupling loss, with a quartic dependence in the limit of $\beta\to1$:

\begin{align}
S(\beta)  
\approx 2\sqrt2\left[1-(1-\beta)^4\right]. 
\end{align}

The complete theory is presented in \SM~Note~\ref{Theory-QD-dynamics}, where the experimental requirements for violating the Bell inequality are also benchmarked in detail (\SM~\figref{fig:Sbenchmark}). 

FIGs.~\ref{fig:1}(a) and (d) show the conceptual scheme of the experiment. We use a narrow linewidth continuous-wave laser as a weak coherent input state $\ket{\alpha}$. The light is coupled in and out of the PhC WG via two mode adaptors with the sample mounted in a cryostat operating at 4~K. Due to high coupling efficiency of the photon-emitter interface, the photon state is strongly modified by the nonlinear interaction, which induces energy-time entanglement involving a continuum of optical modes in multidimensional Hilbert space. The photon pairs generated in different temporal modes can subsequently be analyzed in a Franson interferometer~\cite{Franson1989}. The present experiment is conducted in the regime of weak resonant excitation, i.e., far below saturation threshold of the emitter.

\begin{figure*}[hbtp!]
	\includegraphics[width=0.95\linewidth, trim=0cm 0cm 0cm 0cm,clip]{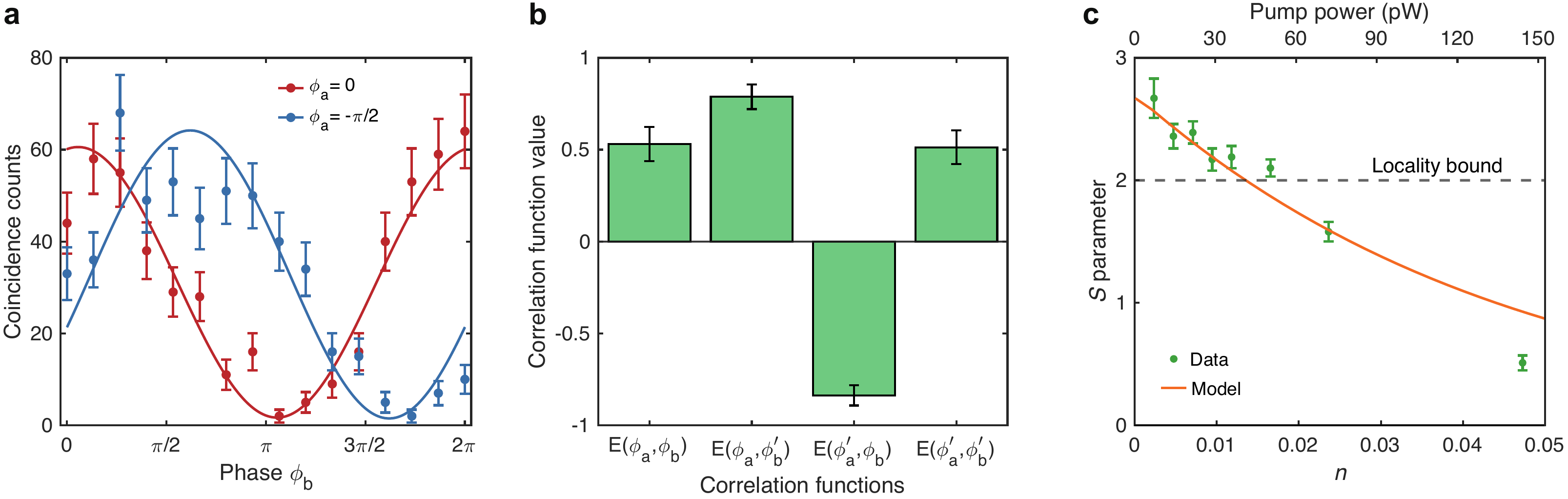}
	\caption{ 
	(color online) \textbf{Two-photon Franson interference measurements  and observation of a violation of the CHSH  inequality.} (a) Interference curves as a function of $\phi_b$ with $\phi_a$ fixed at 0 (red) and $\pi/4$ (blue). The data are modeled with a sinusoidal model whereby a visibility of 95(4)\% is extracted. (b) Measured correlation functions from which $S = 2.67(16)$ is recorded. (c) $S$ parameter versus $n$ (bottom x-axis) or the corresponding pump power in the PhC WG (top x-axis). $n$ and pump powers are calibrated by fitting the full set of transmission intensity data (\SM~Note~\ref{RT}). The solid curve is the theoretical model (\SM~Note~\ref{fit}) with parameters taken from the filtered power saturation $g^{(2)}$ measurements in \SM~Note~\ref{g2}, i.e., no additional fitting was performed. The black dashed line represents the locality bound. The data in (a) and (b) are recorded for the lowest $n$ (0.0024) or pump power (7.2 pW) in (c). The error bars show the standard deviations assuming Poissonian distribution for the photon statistics.}
	\label{fig:3}
\end{figure*}

The quantum correlations induced by the nonlinear scattering are illustrated by the two-photon joint spectral and temporal intensity distributions in FIGs.~\ref{fig:1}(e)-(f). The input weak coherent state  $\ket{\alpha}$ resembles a Dirac delta function in frequency, while the output entangled photon pair $\ket{2_I}$  is Lorentzian broadened by the QD linewidth $\Gamma$. It can be expressed as $\ket{2_I} = \frac{1}{2} \int{d\Delta} \mathcal{T}_{\Delta,-\Delta} \ket{1_{\Delta}}\ket{1_{-\Delta}}$ (\SM~Note~\ref{scattering}), where $\Delta = \Delta_a = -\Delta_b$ is the frequency detuning of each outgoing photon relative to the pump frequency, and $\mathcal{T}_{\Delta,-\Delta} = -4\beta^2/[\pi\Gamma (1 + 4\frac{\Delta^2}{\Gamma^2})]$ is the two-photon Lorentzian spectrum \cite{LeJeannic2021}. Energy conservation demands $2\omega_p = \omega_a + \omega_b$, which introduces anti-correlation in the two-photon joint spectral density, cf. \figref{fig:1}(e). The time uncertainty of the generated  photon pair is determined by the pump laser coherence time $\tau_L > 1~\mu$s (inversely proportional to the laser linewidth $\Gamma_L/2\pi \approx 100$ kHz), which is much longer than the Purcell enhanced QD lifetime $\tau_{QD}\approx$ 69~ps (\figref{fig:1} (f)).

\figref{fig:1}(b) measures the transmission intensity of scattered photons versus QD detuning using one of the two unbalanced Mach-Zehnder interferometers (UMZIs) at different phases $\phi$ (see UMZI setup in \figref{fig:2}(b)). This allows separate measurements of either the extinction of elastically scattered photons from the weak coherent state (blue, $\phi=0$) or the inelastically scattered photons (red, $\phi=\pi$). On resonance, the single-photon component is primarily reflected due to destructive interference in the PhC WG, while the transmitted mode consists of residual coherent photons from the laser and inelastically scattered photons. For $\phi=0$, elastic scattering dominates as the laser photons traversing the short and long paths constructively interfere at the beamsplitter of one UMZI, thereby revealing a transmission dip (blue data in \figref{fig:1}(b)). Conversely, for $\phi=\pi$, these laser photons destructively interfere, allowing direct observation of the inelastically scattered photons (red data in \figref{fig:1}(b)).

The pronounced extinction of the transmission intensity is indicative of efficient radiative coupling to the PhC WG. By modelling the experimental data sets, we extract $\beta =92 \%$ and a Purcell enhancement from slow light in the PhC WG of  $F_P \approx 15.9$, which increases the QD decay rate to $\Gamma/2\pi = 2.3$~GHz (compared to $\approx0.14$~GHz for QDs in bulk~\cite{Arcari2014}). In the entanglement characterization discussed below, a narrow bandwidth notch spectral filter is implemented to suppress residual laser leakage due to experimental imperfections (non-unity $\beta$-factor, residual slow spectral diffusion, etc), see \SM~Note~\ref{filter} and \figref{fig:spectra_params} for further details of the filter and its impact on various parameter estimates of the QD.

The successful preparation of a two-photon component is quantified by second-order photon correlation measurements. Here we observed a pronounced photon bunching of $g^{(2)}(0) \approx 210$ after applying the spectral filter, see~\figref{fig:1}(c). This explicitly demonstrates that the incoming Poissonian photon distribution is significantly altered by strong nonlinear interaction with the QD~\cite{LeJeannic2021,Chang2007}. While our theoretical model does not take the filter into account, it accurately describes the experimental data of \figref{fig:1}(c), by adjusting the input model parameters, see \SM~Notes~\ref{g2} and \ref{fit} for the details and justification. We modelled both the unfiltered and filtered sets of second-order correlation ($g^{(2)}$) data versus the mean number of pump photons $n$, see \SM~Note~\ref{g2}. 

\figref{fig:2}(b) illustrates the experimental setup. The scattered light from the QD PhC WG is spectrally filtered and directed by a nonpolarizing fiber beamsplitter to two identical UMZIs for entanglement analysis. The time difference between the two arms of each UMZI is set to $\tau_I=3.6$~ns, which is shorter than $\tau_L$ and longer than the two-photon correlation time set by $\tau_{QD}$. We implement two-photon Franson interference measurements by recording time-resolved correlations between photon pairs while controlling the interferometric phases ($\phi_a$ and $\phi_b$). \figref{fig:2}(a) reveals three distinct correlation peaks, corresponding to every possible path that the two photons can take separately: long-short $\ket{l,s}$ (left), short-short $\ket{s,s}$ or long-long $\ket{l,l}$ (center), and short-long $\ket{s,l}$ (right). Since the two paths $\ket{s,s}$ and $\ket{l,l}$ cannot be distinguished in time, the central peak corresponds to projection onto the entangled state $\ket{s,s} + e^{i(\phi_a + \phi_b)}\ket{l,l}$. By tuning the interferometer phases such that $\phi_a + \phi_b=0~(\pi)$, we observe constructive (destructive) interference of the central peak (see  \figref{fig:2}(a)), stemming from the two-photon energy-time entanglement. By further measuring a 2D histogram shown in \figref{fig:2}(c), almost background-free quantum interference is observed as a testimony of the highly efficient spectral selection of the two-photon scattering component, see \SM~Note~\ref{Franson} for background noise comparison between filtered and unfiltered data. 

In \figref{fig:3}(a), we scan the phase $\phi_b$ for two different phase settings of interferometer $a$  ($\phi_a=0,\pi/4$). The Franson interference visibility is defined as $V =(R_\text{max} - R_\text{min})/(R_\text{max}+R_\text{min})$, where $R_\text{min}$ ($R_\text{max}$) is the coincidence rate of the central peak at the minimum (maximum) of the interference curve. To obtain higher count rates with smaller fluctuations (error bars), the coincidence time window is increased to 0.512 ns compared with its counterpart (0.064 ns) in FIGs.~\ref{fig:1}-\ref{fig:2}. Fitting the data with a sinusoidal, we extract an interference visibility of $V=95(4)\%$, which indicates the presence of entanglement~\cite{Kwiat1993}. 

The energy-time entangled photon pair induced by the nonlinear interaction is thoroughly certified with a CHSH Bell inequality test \cite{Clauser1969}. The CHSH $S$ parameter is defined as $S= \left|E(\phi_a,\phi_b) + E(\phi_a,\phi_{b'}) - E(\phi_{a'},\phi_b) + E(\phi_{a'},\phi_{b'})\right|$, where $E(\phi_a,\phi_b)$ denotes the correlation function required for the CHSH inequality, which is a combination of four unnormalized $g^{(2)}$ after the UMZIs at different phase settings  (\SM Note~\ref{CHSH}). \figref{fig:3}(b) shows the strongest correlations measured at the lowest value of $n$ in \figref{fig:3}(c), which corresponds to a pump power of 7.2 pW at a single photon level. We record a pronounced violation of the CHSH Bell inequality $S=2.67(16)>2$ by more than four standard deviations.
This validates that non-local quantum correlations can be induced by two-photon inelastic scattering of a deterministically coupled two-level emitter.  \figref{fig:3}(c) explores the power dependence of the $S$ parameter and the experimental data agree well with the theoretical model detailed in \SM~Notes~\ref{fit}. The entanglement quality is primarily limited by photon distinguishability contributions from pure dephasing, as well as multi-photon scattering processes from finite $n$.

We have experimentally demonstrated the violation of the CHSH Bell inequality by weak scattering of a single two-level emitter deterministically coupled to light in a PhC WG. Our scheme operates at a picowatt pump power (single-photon level) by exploiting the giant nonlinearity of the emitter, laying the foundation for ultimately energy-efficient and integrated sources of entangled photons. The source is easy to operate experimentally since no elaborate excitation schemes or active spin control is required, which could constitute significant overhead for future up-scaling of entanglement generation schemes. Future experiments could exploit the creation of high-dimensional entanglement \cite{Martin2017} and the synthesis of photonic quantum states useful for quantum optics neural network \cite{Steinbrecher2019}. Another promising direction is to engineer the inelastic scattering processes by many-body subradiant states using coupled QDs ~\cite{Ke2019a,Tiranov2023}. Waveguide-mediated quantum nonlinear interactions will prove essential to applications within photonic quantum computing~\cite{Alexander2016a}, quantum communication~\cite{Zhong2015} and quantum sensing~\cite{Carlos2021,Kaiser2018}.\\
\newline
\textbf{Data availability}
\newline
The data that support the figures in this manuscript are available from the corresponding authors upon request.
\newline\\
\textbf{Acknowledgement}
\newline
We gratefully acknowledge financial support from Danmarks Grundforskningsfond (DNRF 139, Hy-Q Center for Hybrid Quantum Networks), the Novo Nordisk Foundation (Challenge
project ”Solid-Q”). Furthermore, this project has received funding from the European Union’s Horizon 2020 research and innovation programmes under Grant Agreements No. 824140 (TOCHA, H2020-FETPROACT-01-2018). B.S. acknowledges financial support from Deutsche Forschungsgemeinschaft (DFG, German Research Foundation), grant no. 449674892. O.A.D.S. acknowledges funding from the European Union’s Horizon 2020 research and innovation program under the Marie Skłodowska-Curie grant agreement no. 801199. M.L.C. acknowledges funding from the European Union’s Horizon 2020 Research and Innovation programme under
grant agreement No. 861097 (project name QUDOT-TECH).
\newline\\
\textbf{Author contributions}
\newline
S.L. and A.T. performed the experiments based on interferometers built by R.B.N., R.G.L. and Y.A. S.L. and A.T. carried out the measurements, and analyzed the data with help from M.L.C. S.L. prepared the figures with input from A.T., A.S.S. and P.L. O.A.D.S. developed the theory with input from B.S and A.S.S. S.L., A.T., O.A.D.S, and P.L. wrote the manuscript with input from all the authors. S.S., A.D.W., and A.L. prepared the wafer. Y.W. and L.M. fabricated the chip. A.T. and P.L. conceived the idea and supervised the project.
\newline\\
\textbf{Competing interests}
\newline
P.L. is the founder of the company
Sparrow Quantum, which commercializes single-photon sources.
The authors declare no other competing interests.

\bibliographystyle{apsrev4-1} 
\let\oldaddcontentsline\addcontentsline
\renewcommand{\addcontentsline}[3]{}
\bibliography{reflist}

\begin{thebibliography}{43}%
\makeatletter
\providecommand \@ifxundefined [1]{%
 \@ifx{#1\undefined}
}%
\providecommand \@ifnum [1]{%
 \ifnum #1\expandafter \@firstoftwo
 \else \expandafter \@secondoftwo
 \fi
}%
\providecommand \@ifx [1]{%
 \ifx #1\expandafter \@firstoftwo
 \else \expandafter \@secondoftwo
 \fi
}%
\providecommand \natexlab [1]{#1}%
\providecommand \enquote  [1]{``#1''}%
\providecommand \bibnamefont  [1]{#1}%
\providecommand \bibfnamefont [1]{#1}%
\providecommand \citenamefont [1]{#1}%
\providecommand \href@noop [0]{\@secondoftwo}%
\providecommand \href [0]{\begingroup \@sanitize@url \@href}%
\providecommand \@href[1]{\@@startlink{#1}\@@href}%
\providecommand \@@href[1]{\endgroup#1\@@endlink}%
\providecommand \@sanitize@url [0]{\catcode `\\12\catcode `\$12\catcode
  `\&12\catcode `\#12\catcode `\^12\catcode `\_12\catcode `\%12\relax}%
\providecommand \@@startlink[1]{}%
\providecommand \@@endlink[0]{}%
\providecommand \url  [0]{\begingroup\@sanitize@url \@url }%
\providecommand \@url [1]{\endgroup\@href {#1}{\urlprefix }}%
\providecommand \urlprefix  [0]{URL }%
\providecommand \Eprint [0]{\href }%
\providecommand \doibase [0]{http://dx.doi.org/}%
\providecommand \selectlanguage [0]{\@gobble}%
\providecommand \bibinfo  [0]{\@secondoftwo}%
\providecommand \bibfield  [0]{\@secondoftwo}%
\providecommand \translation [1]{[#1]}%
\providecommand \BibitemOpen [0]{}%
\providecommand \bibitemStop [0]{}%
\providecommand \bibitemNoStop [0]{.\EOS\space}%
\providecommand \EOS [0]{\spacefactor3000\relax}%
\providecommand \BibitemShut  [1]{\csname bibitem#1\endcsname}%
\let\auto@bib@innerbib\@empty
\bibitem [{\citenamefont {Shen}\ and\ \citenamefont {Fan}(2007)}]{Shen2007}%
  \BibitemOpen
  \bibfield  {author} {\bibinfo {author} {\bibfnamefont {J.-T.}\ \bibnamefont
  {Shen}}\ and\ \bibinfo {author} {\bibfnamefont {S.}~\bibnamefont {Fan}},\
  }\href {\doibase 10.1103/PhysRevLett.98.153003} {\bibfield  {journal}
  {\bibinfo  {journal} {Phys. Rev. Lett.}\ }\textbf {\bibinfo {volume} {98}},\
  \bibinfo {pages} {153003} (\bibinfo {year} {2007})}\BibitemShut {NoStop}%
\bibitem [{\citenamefont {Kiilerich}\ and\ \citenamefont
  {M\o{}lmer}(2019)}]{Kiilerich2019}%
  \BibitemOpen
  \bibfield  {author} {\bibinfo {author} {\bibfnamefont {A.~H.}\ \bibnamefont
  {Kiilerich}}\ and\ \bibinfo {author} {\bibfnamefont {K.}~\bibnamefont
  {M\o{}lmer}},\ }\href {\doibase 10.1103/PhysRevLett.123.123604} {\bibfield
  {journal} {\bibinfo  {journal} {Phys. Rev. Lett.}\ }\textbf {\bibinfo
  {volume} {123}},\ \bibinfo {pages} {123604} (\bibinfo {year}
  {2019})}\BibitemShut {NoStop}%
\bibitem [{\citenamefont {Lodahl}\ \emph {et~al.}(2022)\citenamefont {Lodahl},
  \citenamefont {Ludwig},\ and\ \citenamefont {Warburton}}]{Lodahl2022}%
  \BibitemOpen
  \bibfield  {author} {\bibinfo {author} {\bibfnamefont {P.}~\bibnamefont
  {Lodahl}}, \bibinfo {author} {\bibfnamefont {A.}~\bibnamefont {Ludwig}}, \
  and\ \bibinfo {author} {\bibfnamefont {R.~J.}\ \bibnamefont {Warburton}},\
  }\href {\doibase 10.1063/PT.3.4962} {\bibfield  {journal} {\bibinfo
  {journal} {Physics Today}\ }\textbf {\bibinfo {volume} {75}},\ \bibinfo
  {pages} {44} (\bibinfo {year} {2022})}\BibitemShut {NoStop}%
\bibitem [{\citenamefont {Javadi}\ \emph {et~al.}(2015)\citenamefont {Javadi},
  \citenamefont {S{\"o}llner}, \citenamefont {Arcari}, \citenamefont {Hansen},
  \citenamefont {Midolo}, \citenamefont {Mahmoodian}, \citenamefont {Kir{\v
  s}ansk{\.e}}, \citenamefont {Pregnolato}, \citenamefont {Lee}, \citenamefont
  {Song}, \citenamefont {Stobbe},\ and\ \citenamefont {Lodahl}}]{Javadi2015a}%
  \BibitemOpen
  \bibfield  {author} {\bibinfo {author} {\bibfnamefont {A.}~\bibnamefont
  {Javadi}}, \bibinfo {author} {\bibfnamefont {I.}~\bibnamefont {S{\"o}llner}},
  \bibinfo {author} {\bibfnamefont {M.}~\bibnamefont {Arcari}}, \bibinfo
  {author} {\bibfnamefont {S.~L.}\ \bibnamefont {Hansen}}, \bibinfo {author}
  {\bibfnamefont {L.}~\bibnamefont {Midolo}}, \bibinfo {author} {\bibfnamefont
  {S.}~\bibnamefont {Mahmoodian}}, \bibinfo {author} {\bibfnamefont
  {G.}~\bibnamefont {Kir{\v s}ansk{\.e}}}, \bibinfo {author} {\bibfnamefont
  {T.}~\bibnamefont {Pregnolato}}, \bibinfo {author} {\bibfnamefont {E.~H.}\
  \bibnamefont {Lee}}, \bibinfo {author} {\bibfnamefont {J.~D.}\ \bibnamefont
  {Song}}, \bibinfo {author} {\bibfnamefont {S.}~\bibnamefont {Stobbe}}, \ and\
  \bibinfo {author} {\bibfnamefont {P.}~\bibnamefont {Lodahl}},\ }\href
  {\doibase 10.1038/ncomms9655} {\bibfield  {journal} {\bibinfo  {journal}
  {Nature Communications}\ }\textbf {\bibinfo {volume} {6}},\ \bibinfo {pages}
  {8655} (\bibinfo {year} {2015})}\BibitemShut {NoStop}%
\bibitem [{\citenamefont {Sipahigil}\ \emph {et~al.}(2016)\citenamefont
  {Sipahigil}, \citenamefont {Evans}, \citenamefont {Sukachev}, \citenamefont
  {Burek}, \citenamefont {Borregaard}, \citenamefont {Bhaskar}, \citenamefont
  {Nguyen}, \citenamefont {Pacheco}, \citenamefont {Atikian}, \citenamefont
  {Meuwly}, \citenamefont {Camacho}, \citenamefont {Jelezko}, \citenamefont
  {Bielejec}, \citenamefont {Park}, \citenamefont {Lon{\v c}ar},\ and\
  \citenamefont {Lukin}}]{Sipahigil2016}%
  \BibitemOpen
  \bibfield  {author} {\bibinfo {author} {\bibfnamefont {A.}~\bibnamefont
  {Sipahigil}}, \bibinfo {author} {\bibfnamefont {R.~E.}\ \bibnamefont
  {Evans}}, \bibinfo {author} {\bibfnamefont {D.~D.}\ \bibnamefont {Sukachev}},
  \bibinfo {author} {\bibfnamefont {M.~J.}\ \bibnamefont {Burek}}, \bibinfo
  {author} {\bibfnamefont {J.}~\bibnamefont {Borregaard}}, \bibinfo {author}
  {\bibfnamefont {M.~K.}\ \bibnamefont {Bhaskar}}, \bibinfo {author}
  {\bibfnamefont {C.~T.}\ \bibnamefont {Nguyen}}, \bibinfo {author}
  {\bibfnamefont {J.~L.}\ \bibnamefont {Pacheco}}, \bibinfo {author}
  {\bibfnamefont {H.~A.}\ \bibnamefont {Atikian}}, \bibinfo {author}
  {\bibfnamefont {C.}~\bibnamefont {Meuwly}}, \bibinfo {author} {\bibfnamefont
  {R.~M.}\ \bibnamefont {Camacho}}, \bibinfo {author} {\bibfnamefont
  {F.}~\bibnamefont {Jelezko}}, \bibinfo {author} {\bibfnamefont
  {E.}~\bibnamefont {Bielejec}}, \bibinfo {author} {\bibfnamefont
  {H.}~\bibnamefont {Park}}, \bibinfo {author} {\bibfnamefont {M.}~\bibnamefont
  {Lon{\v c}ar}}, \ and\ \bibinfo {author} {\bibfnamefont {M.~D.}\ \bibnamefont
  {Lukin}},\ }\href {\doibase 10.1126/science.aah6875} {\bibfield  {journal}
  {\bibinfo  {journal} {Science}\ }\textbf {\bibinfo {volume} {354}},\ \bibinfo
  {pages} {847} (\bibinfo {year} {2016})}\BibitemShut {NoStop}%
\bibitem [{\citenamefont {Hamsen}\ \emph {et~al.}(2017)\citenamefont {Hamsen},
  \citenamefont {Tolazzi}, \citenamefont {Wilk},\ and\ \citenamefont
  {Rempe}}]{Hamsen2017}%
  \BibitemOpen
  \bibfield  {author} {\bibinfo {author} {\bibfnamefont {C.}~\bibnamefont
  {Hamsen}}, \bibinfo {author} {\bibfnamefont {K.~N.}\ \bibnamefont {Tolazzi}},
  \bibinfo {author} {\bibfnamefont {T.}~\bibnamefont {Wilk}}, \ and\ \bibinfo
  {author} {\bibfnamefont {G.}~\bibnamefont {Rempe}},\ }\href {\doibase
  10.1103/PhysRevLett.118.133604} {\bibfield  {journal} {\bibinfo  {journal}
  {Phys. Rev. Lett.}\ }\textbf {\bibinfo {volume} {118}},\ \bibinfo {pages}
  {133604} (\bibinfo {year} {2017})}\BibitemShut {NoStop}%
\bibitem [{\citenamefont {Pscherer}\ \emph {et~al.}(2021)\citenamefont
  {Pscherer}, \citenamefont {Meierhofer}, \citenamefont {Wang}, \citenamefont
  {Kelkar}, \citenamefont {Mart\'{\i}n-Cano}, \citenamefont {Utikal},
  \citenamefont {G\"otzinger},\ and\ \citenamefont
  {Sandoghdar}}]{Pscherer2021}%
  \BibitemOpen
  \bibfield  {author} {\bibinfo {author} {\bibfnamefont {A.}~\bibnamefont
  {Pscherer}}, \bibinfo {author} {\bibfnamefont {M.}~\bibnamefont
  {Meierhofer}}, \bibinfo {author} {\bibfnamefont {D.}~\bibnamefont {Wang}},
  \bibinfo {author} {\bibfnamefont {H.}~\bibnamefont {Kelkar}}, \bibinfo
  {author} {\bibfnamefont {D.}~\bibnamefont {Mart\'{\i}n-Cano}}, \bibinfo
  {author} {\bibfnamefont {T.}~\bibnamefont {Utikal}}, \bibinfo {author}
  {\bibfnamefont {S.}~\bibnamefont {G\"otzinger}}, \ and\ \bibinfo {author}
  {\bibfnamefont {V.}~\bibnamefont {Sandoghdar}},\ }\href {\doibase
  10.1103/PhysRevLett.127.133603} {\bibfield  {journal} {\bibinfo  {journal}
  {Phys. Rev. Lett.}\ }\textbf {\bibinfo {volume} {127}},\ \bibinfo {pages}
  {133603} (\bibinfo {year} {2021})}\BibitemShut {NoStop}%
\bibitem [{\citenamefont {Schulte}\ \emph {et~al.}(2015)\citenamefont
  {Schulte}, \citenamefont {Hansom}, \citenamefont {Jones}, \citenamefont
  {Matthiesen}, \citenamefont {Le~Gall},\ and\ \citenamefont
  {Atat{\"u}re}}]{Schulte2015}%
  \BibitemOpen
  \bibfield  {author} {\bibinfo {author} {\bibfnamefont {C.~H.~H.}\
  \bibnamefont {Schulte}}, \bibinfo {author} {\bibfnamefont {J.}~\bibnamefont
  {Hansom}}, \bibinfo {author} {\bibfnamefont {A.~E.}\ \bibnamefont {Jones}},
  \bibinfo {author} {\bibfnamefont {C.}~\bibnamefont {Matthiesen}}, \bibinfo
  {author} {\bibfnamefont {C.}~\bibnamefont {Le~Gall}}, \ and\ \bibinfo
  {author} {\bibfnamefont {M.}~\bibnamefont {Atat{\"u}re}},\ }\href {\doibase
  10.1038/nature14868} {\bibfield  {journal} {\bibinfo  {journal} {Nature}\
  }\textbf {\bibinfo {volume} {525}},\ \bibinfo {pages} {222} (\bibinfo {year}
  {2015})}\BibitemShut {NoStop}%
\bibitem [{\citenamefont {Hinney}\ \emph {et~al.}(2021)\citenamefont {Hinney},
  \citenamefont {Prasad}, \citenamefont {Mahmoodian}, \citenamefont {Hammerer},
  \citenamefont {Rauschenbeutel}, \citenamefont {Schneeweiss}, \citenamefont
  {Volz},\ and\ \citenamefont {Schemmer}}]{Hinney2021}%
  \BibitemOpen
  \bibfield  {author} {\bibinfo {author} {\bibfnamefont {J.}~\bibnamefont
  {Hinney}}, \bibinfo {author} {\bibfnamefont {A.~S.}\ \bibnamefont {Prasad}},
  \bibinfo {author} {\bibfnamefont {S.}~\bibnamefont {Mahmoodian}}, \bibinfo
  {author} {\bibfnamefont {K.}~\bibnamefont {Hammerer}}, \bibinfo {author}
  {\bibfnamefont {A.}~\bibnamefont {Rauschenbeutel}}, \bibinfo {author}
  {\bibfnamefont {P.}~\bibnamefont {Schneeweiss}}, \bibinfo {author}
  {\bibfnamefont {J.}~\bibnamefont {Volz}}, \ and\ \bibinfo {author}
  {\bibfnamefont {M.}~\bibnamefont {Schemmer}},\ }\href {\doibase
  10.1103/PhysRevLett.127.123602} {\bibfield  {journal} {\bibinfo  {journal}
  {Phys. Rev. Lett.}\ }\textbf {\bibinfo {volume} {127}},\ \bibinfo {pages}
  {123602} (\bibinfo {year} {2021})}\BibitemShut {NoStop}%
\bibitem [{\citenamefont {Jeannic}\ \emph {et~al.}(2022)\citenamefont
  {Jeannic}, \citenamefont {Tiranov}, \citenamefont {Carolan}, \citenamefont
  {Ramos}, \citenamefont {Wang}, \citenamefont {Appel}, \citenamefont {Scholz},
  \citenamefont {Wieck}, \citenamefont {Ludwig}, \citenamefont {Rotenberg},
  \citenamefont {Midolo}, \citenamefont {{Garc{\'i}a-Ripoll}}, \citenamefont
  {S{\o}rensen},\ and\ \citenamefont {Lodahl}}]{Jeannic2022}%
  \BibitemOpen
  \bibfield  {author} {\bibinfo {author} {\bibfnamefont {H.~L.}\ \bibnamefont
  {Jeannic}}, \bibinfo {author} {\bibfnamefont {A.}~\bibnamefont {Tiranov}},
  \bibinfo {author} {\bibfnamefont {J.}~\bibnamefont {Carolan}}, \bibinfo
  {author} {\bibfnamefont {T.}~\bibnamefont {Ramos}}, \bibinfo {author}
  {\bibfnamefont {Y.}~\bibnamefont {Wang}}, \bibinfo {author} {\bibfnamefont
  {M.~H.}\ \bibnamefont {Appel}}, \bibinfo {author} {\bibfnamefont
  {S.}~\bibnamefont {Scholz}}, \bibinfo {author} {\bibfnamefont {A.~D.}\
  \bibnamefont {Wieck}}, \bibinfo {author} {\bibfnamefont {A.}~\bibnamefont
  {Ludwig}}, \bibinfo {author} {\bibfnamefont {N.}~\bibnamefont {Rotenberg}},
  \bibinfo {author} {\bibfnamefont {L.}~\bibnamefont {Midolo}}, \bibinfo
  {author} {\bibfnamefont {J.~J.}\ \bibnamefont {{Garc{\'i}a-Ripoll}}},
  \bibinfo {author} {\bibfnamefont {A.~S.}\ \bibnamefont {S{\o}rensen}}, \ and\
  \bibinfo {author} {\bibfnamefont {P.}~\bibnamefont {Lodahl}},\ }\href
  {\doibase 10.1038/s41567-022-01720-x} {\bibfield  {journal} {\bibinfo
  {journal} {Nature Physics}\ }\textbf {\bibinfo {volume} {18}},\ \bibinfo
  {pages} {1191} (\bibinfo {year} {2022})}\BibitemShut {NoStop}%
\bibitem [{\citenamefont {Tomm}\ \emph {et~al.}(2023)\citenamefont {Tomm},
  \citenamefont {Mahmoodian}, \citenamefont {Antoniadis}, \citenamefont
  {Schott}, \citenamefont {Valentin}, \citenamefont {Wieck}, \citenamefont
  {Ludwig}, \citenamefont {Javadi},\ and\ \citenamefont
  {Warburton}}]{Tomm2023}%
  \BibitemOpen
  \bibfield  {author} {\bibinfo {author} {\bibfnamefont {N.}~\bibnamefont
  {Tomm}}, \bibinfo {author} {\bibfnamefont {S.}~\bibnamefont {Mahmoodian}},
  \bibinfo {author} {\bibfnamefont {N.~O.}\ \bibnamefont {Antoniadis}},
  \bibinfo {author} {\bibfnamefont {R.}~\bibnamefont {Schott}}, \bibinfo
  {author} {\bibfnamefont {S.~R.}\ \bibnamefont {Valentin}}, \bibinfo {author}
  {\bibfnamefont {A.~D.}\ \bibnamefont {Wieck}}, \bibinfo {author}
  {\bibfnamefont {A.}~\bibnamefont {Ludwig}}, \bibinfo {author} {\bibfnamefont
  {A.}~\bibnamefont {Javadi}}, \ and\ \bibinfo {author} {\bibfnamefont {R.~J.}\
  \bibnamefont {Warburton}},\ }\href {\doibase 10.1038/s41567-023-01997-6}
  {\bibfield  {journal} {\bibinfo  {journal} {Nature Physics}\ } (\bibinfo
  {year} {2023}),\ 10.1038/s41567-023-01997-6}\BibitemShut {NoStop}%
\bibitem [{\citenamefont {Witthaut}\ \emph {et~al.}(2012)\citenamefont
  {Witthaut}, \citenamefont {Lukin},\ and\ \citenamefont
  {Sørensen}}]{Witthaut2012}%
  \BibitemOpen
  \bibfield  {author} {\bibinfo {author} {\bibfnamefont {D.}~\bibnamefont
  {Witthaut}}, \bibinfo {author} {\bibfnamefont {M.~D.}\ \bibnamefont {Lukin}},
  \ and\ \bibinfo {author} {\bibfnamefont {A.~S.}\ \bibnamefont {Sørensen}},\
  }\href {\doibase 10.1209/0295-5075/97/50007} {\bibfield  {journal} {\bibinfo
  {journal} {Europhysics Letters}\ }\textbf {\bibinfo {volume} {97}},\ \bibinfo
  {pages} {50007} (\bibinfo {year} {2012})}\BibitemShut {NoStop}%
\bibitem [{\citenamefont {Konyk}\ and\ \citenamefont
  {{Gea-Banacloche}}(2019)}]{Konyk2019}%
  \BibitemOpen
  \bibfield  {author} {\bibinfo {author} {\bibfnamefont {W.}~\bibnamefont
  {Konyk}}\ and\ \bibinfo {author} {\bibfnamefont {J.}~\bibnamefont
  {{Gea-Banacloche}}},\ }\href {\doibase 10.1103/PhysRevA.99.010301} {\bibfield
   {journal} {\bibinfo  {journal} {Physical Review A}\ }\textbf {\bibinfo
  {volume} {99}},\ \bibinfo {pages} {010301} (\bibinfo {year}
  {2019})}\BibitemShut {NoStop}%
\bibitem [{\citenamefont {Krastanov}\ \emph {et~al.}(2022)\citenamefont
  {Krastanov}, \citenamefont {Jacobs}, \citenamefont {Gilbert}, \citenamefont
  {Englund},\ and\ \citenamefont {Heuck}}]{Krastanov2022}%
  \BibitemOpen
  \bibfield  {author} {\bibinfo {author} {\bibfnamefont {S.}~\bibnamefont
  {Krastanov}}, \bibinfo {author} {\bibfnamefont {K.}~\bibnamefont {Jacobs}},
  \bibinfo {author} {\bibfnamefont {G.}~\bibnamefont {Gilbert}}, \bibinfo
  {author} {\bibfnamefont {D.~R.}\ \bibnamefont {Englund}}, \ and\ \bibinfo
  {author} {\bibfnamefont {M.}~\bibnamefont {Heuck}},\ }\href {\doibase
  10.1038/s41534-022-00604-5} {\bibfield  {journal} {\bibinfo  {journal} {npj
  Quantum Information}\ }\textbf {\bibinfo {volume} {8}},\ \bibinfo {pages}
  {103} (\bibinfo {year} {2022})}\BibitemShut {NoStop}%
\bibitem [{\citenamefont {Chang}\ \emph {et~al.}(2007)\citenamefont {Chang},
  \citenamefont {S{\o}rensen}, \citenamefont {Demler},\ and\ \citenamefont
  {Lukin}}]{Chang2007}%
  \BibitemOpen
  \bibfield  {author} {\bibinfo {author} {\bibfnamefont {D.~E.}\ \bibnamefont
  {Chang}}, \bibinfo {author} {\bibfnamefont {A.~S.}\ \bibnamefont
  {S{\o}rensen}}, \bibinfo {author} {\bibfnamefont {E.~A.}\ \bibnamefont
  {Demler}}, \ and\ \bibinfo {author} {\bibfnamefont {M.~D.}\ \bibnamefont
  {Lukin}},\ }\href {\doibase 10.1038/nphys708} {\bibfield  {journal} {\bibinfo
   {journal} {Nature Physics}\ }\textbf {\bibinfo {volume} {3}},\ \bibinfo
  {pages} {807} (\bibinfo {year} {2007})}\BibitemShut {NoStop}%
\bibitem [{\citenamefont {Fayard}\ \emph {et~al.}(2021)\citenamefont {Fayard},
  \citenamefont {Henriet}, \citenamefont {{Asenjo-Garcia}},\ and\ \citenamefont
  {Chang}}]{Fayard2021}%
  \BibitemOpen
  \bibfield  {author} {\bibinfo {author} {\bibfnamefont {N.}~\bibnamefont
  {Fayard}}, \bibinfo {author} {\bibfnamefont {L.}~\bibnamefont {Henriet}},
  \bibinfo {author} {\bibfnamefont {A.}~\bibnamefont {{Asenjo-Garcia}}}, \ and\
  \bibinfo {author} {\bibfnamefont {D.~E.}\ \bibnamefont {Chang}},\ }\href
  {\doibase 10.1103/PhysRevResearch.3.033233} {\bibfield  {journal} {\bibinfo
  {journal} {Physical Review Research}\ }\textbf {\bibinfo {volume} {3}},\
  \bibinfo {pages} {033233} (\bibinfo {year} {2021})}\BibitemShut {NoStop}%
\bibitem [{\citenamefont {Bello}\ \emph {et~al.}(2022)\citenamefont {Bello},
  \citenamefont {Platero},\ and\ \citenamefont
  {{Gonz{\'a}lez-Tudela}}}]{Bello2022}%
  \BibitemOpen
  \bibfield  {author} {\bibinfo {author} {\bibfnamefont {M.}~\bibnamefont
  {Bello}}, \bibinfo {author} {\bibfnamefont {G.}~\bibnamefont {Platero}}, \
  and\ \bibinfo {author} {\bibfnamefont {A.}~\bibnamefont
  {{Gonz{\'a}lez-Tudela}}},\ }\href {\doibase 10.1103/PRXQuantum.3.010336}
  {\bibfield  {journal} {\bibinfo  {journal} {PRX Quantum}\ }\textbf {\bibinfo
  {volume} {3}},\ \bibinfo {pages} {010336} (\bibinfo {year}
  {2022})}\BibitemShut {NoStop}%
\bibitem [{\citenamefont {Bourassa}\ \emph {et~al.}(2021)\citenamefont
  {Bourassa}, \citenamefont {Alexander}, \citenamefont {Vasmer}, \citenamefont
  {Patil}, \citenamefont {Tzitrin}, \citenamefont {Matsuura}, \citenamefont
  {Su}, \citenamefont {Baragiola}, \citenamefont {Guha}, \citenamefont
  {Dauphinais}, \citenamefont {Sabapathy}, \citenamefont {Menicucci},\ and\
  \citenamefont {Dhand}}]{Bourassa2021}%
  \BibitemOpen
  \bibfield  {author} {\bibinfo {author} {\bibfnamefont {J.~E.}\ \bibnamefont
  {Bourassa}}, \bibinfo {author} {\bibfnamefont {R.~N.}\ \bibnamefont
  {Alexander}}, \bibinfo {author} {\bibfnamefont {M.}~\bibnamefont {Vasmer}},
  \bibinfo {author} {\bibfnamefont {A.}~\bibnamefont {Patil}}, \bibinfo
  {author} {\bibfnamefont {I.}~\bibnamefont {Tzitrin}}, \bibinfo {author}
  {\bibfnamefont {T.}~\bibnamefont {Matsuura}}, \bibinfo {author}
  {\bibfnamefont {D.}~\bibnamefont {Su}}, \bibinfo {author} {\bibfnamefont
  {B.~Q.}\ \bibnamefont {Baragiola}}, \bibinfo {author} {\bibfnamefont
  {S.}~\bibnamefont {Guha}}, \bibinfo {author} {\bibfnamefont {G.}~\bibnamefont
  {Dauphinais}}, \bibinfo {author} {\bibfnamefont {K.~K.}\ \bibnamefont
  {Sabapathy}}, \bibinfo {author} {\bibfnamefont {N.~C.}\ \bibnamefont
  {Menicucci}}, \ and\ \bibinfo {author} {\bibfnamefont {I.}~\bibnamefont
  {Dhand}},\ }\href {\doibase 10.22331/q-2021-02-04-392} {\bibfield  {journal}
  {\bibinfo  {journal} {Quantum}\ }\textbf {\bibinfo {volume} {5}},\ \bibinfo
  {pages} {392} (\bibinfo {year} {2021})},\ \Eprint
  {http://arxiv.org/abs/2010.02905} {arxiv:2010.02905 [quant-ph]} \BibitemShut
  {NoStop}%
\bibitem [{\citenamefont {Peiris}\ \emph {et~al.}(2017)\citenamefont {Peiris},
  \citenamefont {Konthasinghe},\ and\ \citenamefont {Muller}}]{Peiris2017a}%
  \BibitemOpen
  \bibfield  {author} {\bibinfo {author} {\bibfnamefont {M.}~\bibnamefont
  {Peiris}}, \bibinfo {author} {\bibfnamefont {K.}~\bibnamefont
  {Konthasinghe}}, \ and\ \bibinfo {author} {\bibfnamefont {A.}~\bibnamefont
  {Muller}},\ }\href {\doibase 10.1103/PhysRevLett.118.030501} {\bibfield
  {journal} {\bibinfo  {journal} {Physical Review Letters}\ }\textbf {\bibinfo
  {volume} {118}},\ \bibinfo {pages} {030501} (\bibinfo {year}
  {2017})}\BibitemShut {NoStop}%
\bibitem [{\citenamefont {Muller}\ \emph {et~al.}(2008)\citenamefont {Muller},
  \citenamefont {Fang}, \citenamefont {Lawall},\ and\ \citenamefont
  {Solomon}}]{Muller2008}%
  \BibitemOpen
  \bibfield  {author} {\bibinfo {author} {\bibfnamefont {A.}~\bibnamefont
  {Muller}}, \bibinfo {author} {\bibfnamefont {W.}~\bibnamefont {Fang}},
  \bibinfo {author} {\bibfnamefont {J.}~\bibnamefont {Lawall}}, \ and\ \bibinfo
  {author} {\bibfnamefont {G.~S.}\ \bibnamefont {Solomon}},\ }\href {\doibase
  10.1103/PhysRevLett.101.027401} {\bibfield  {journal} {\bibinfo  {journal}
  {Physical Review Letters}\ }\textbf {\bibinfo {volume} {101}},\ \bibinfo
  {pages} {027401} (\bibinfo {year} {2008})}\BibitemShut {NoStop}%
\bibitem [{\citenamefont {Jayakumar}\ \emph {et~al.}(2013)\citenamefont
  {Jayakumar}, \citenamefont {Predojevi{\'c}}, \citenamefont {Huber},
  \citenamefont {Kauten}, \citenamefont {Solomon},\ and\ \citenamefont
  {Weihs}}]{Jayakumar2013}%
  \BibitemOpen
  \bibfield  {author} {\bibinfo {author} {\bibfnamefont {H.}~\bibnamefont
  {Jayakumar}}, \bibinfo {author} {\bibfnamefont {A.}~\bibnamefont
  {Predojevi{\'c}}}, \bibinfo {author} {\bibfnamefont {T.}~\bibnamefont
  {Huber}}, \bibinfo {author} {\bibfnamefont {T.}~\bibnamefont {Kauten}},
  \bibinfo {author} {\bibfnamefont {G.~S.}\ \bibnamefont {Solomon}}, \ and\
  \bibinfo {author} {\bibfnamefont {G.}~\bibnamefont {Weihs}},\ }\href
  {\doibase 10.1103/PhysRevLett.110.135505} {\bibfield  {journal} {\bibinfo
  {journal} {Physical Review Letters}\ }\textbf {\bibinfo {volume} {110}},\
  \bibinfo {pages} {135505} (\bibinfo {year} {2013})}\BibitemShut {NoStop}%
\bibitem [{\citenamefont {Hohn}\ \emph {et~al.}(2023)\citenamefont {Hohn},
  \citenamefont {Barkemeyer}, \citenamefont {von Helversen}, \citenamefont
  {Bremer}, \citenamefont {Gschrey}, \citenamefont {Schulze}, \citenamefont
  {Strittmatter}, \citenamefont {Carmele}, \citenamefont {Rodt}, \citenamefont
  {Bounouar},\ and\ \citenamefont {Reitzenstein}}]{Hohn2023}%
  \BibitemOpen
  \bibfield  {author} {\bibinfo {author} {\bibfnamefont {M.}~\bibnamefont
  {Hohn}}, \bibinfo {author} {\bibfnamefont {K.}~\bibnamefont {Barkemeyer}},
  \bibinfo {author} {\bibfnamefont {M.}~\bibnamefont {von Helversen}}, \bibinfo
  {author} {\bibfnamefont {L.}~\bibnamefont {Bremer}}, \bibinfo {author}
  {\bibfnamefont {M.}~\bibnamefont {Gschrey}}, \bibinfo {author} {\bibfnamefont
  {J.~H.}\ \bibnamefont {Schulze}}, \bibinfo {author} {\bibfnamefont
  {A.}~\bibnamefont {Strittmatter}}, \bibinfo {author} {\bibfnamefont
  {A.}~\bibnamefont {Carmele}}, \bibinfo {author} {\bibfnamefont
  {S.}~\bibnamefont {Rodt}}, \bibinfo {author} {\bibfnamefont {S.}~\bibnamefont
  {Bounouar}}, \ and\ \bibinfo {author} {\bibfnamefont {S.}~\bibnamefont
  {Reitzenstein}},\ }\href {\doibase 10.48550/ARXIV.2301.05697} {\bibfield
  {journal} {\bibinfo  {journal} {arXiv:}\ }\textbf {\bibinfo {volume}
  {2301.05697}} (\bibinfo {year} {2023}),\
  10.48550/ARXIV.2301.05697}\BibitemShut {NoStop}%
\bibitem [{\citenamefont {Glorieux}\ \emph {et~al.}(2010)\citenamefont
  {Glorieux}, \citenamefont {Dubessy}, \citenamefont {Guibal}, \citenamefont
  {Guidoni}, \citenamefont {Likforman}, \citenamefont {Coudreau},\ and\
  \citenamefont {Arimondo}}]{Glorieux2010}%
  \BibitemOpen
  \bibfield  {author} {\bibinfo {author} {\bibfnamefont {Q.}~\bibnamefont
  {Glorieux}}, \bibinfo {author} {\bibfnamefont {R.}~\bibnamefont {Dubessy}},
  \bibinfo {author} {\bibfnamefont {S.}~\bibnamefont {Guibal}}, \bibinfo
  {author} {\bibfnamefont {L.}~\bibnamefont {Guidoni}}, \bibinfo {author}
  {\bibfnamefont {J.-P.}\ \bibnamefont {Likforman}}, \bibinfo {author}
  {\bibfnamefont {T.}~\bibnamefont {Coudreau}}, \ and\ \bibinfo {author}
  {\bibfnamefont {E.}~\bibnamefont {Arimondo}},\ }\href {\doibase
  10.1103/PhysRevA.82.033819} {\bibfield  {journal} {\bibinfo  {journal}
  {Physical Review A}\ }\textbf {\bibinfo {volume} {82}},\ \bibinfo {pages}
  {033819} (\bibinfo {year} {2010})}\BibitemShut {NoStop}%
\bibitem [{\citenamefont {Lodahl}\ \emph {et~al.}(2015)\citenamefont {Lodahl},
  \citenamefont {Mahmoodian},\ and\ \citenamefont {Stobbe}}]{Lodahl2015}%
  \BibitemOpen
  \bibfield  {author} {\bibinfo {author} {\bibfnamefont {P.}~\bibnamefont
  {Lodahl}}, \bibinfo {author} {\bibfnamefont {S.}~\bibnamefont {Mahmoodian}},
  \ and\ \bibinfo {author} {\bibfnamefont {S.}~\bibnamefont {Stobbe}},\ }\href
  {\doibase 10.1103/RevModPhys.87.347} {\bibfield  {journal} {\bibinfo
  {journal} {Reviews of Modern Physics}\ }\textbf {\bibinfo {volume} {87}},\
  \bibinfo {pages} {347} (\bibinfo {year} {2015})}\BibitemShut {NoStop}%
\bibitem [{\citenamefont {Franson}(1989)}]{Franson1989}%
  \BibitemOpen
  \bibfield  {author} {\bibinfo {author} {\bibfnamefont {J.~D.}\ \bibnamefont
  {Franson}},\ }\href {\doibase 10.1103/PhysRevLett.62.2205} {\bibfield
  {journal} {\bibinfo  {journal} {Phys. Rev. Lett.}\ }\textbf {\bibinfo
  {volume} {62}},\ \bibinfo {pages} {2205} (\bibinfo {year}
  {1989})}\BibitemShut {NoStop}%
\bibitem [{\citenamefont {Clauser}\ \emph {et~al.}(1969)\citenamefont
  {Clauser}, \citenamefont {Horne}, \citenamefont {Shimony},\ and\
  \citenamefont {Holt}}]{Clauser1969}%
  \BibitemOpen
  \bibfield  {author} {\bibinfo {author} {\bibfnamefont {J.~F.}\ \bibnamefont
  {Clauser}}, \bibinfo {author} {\bibfnamefont {M.~A.}\ \bibnamefont {Horne}},
  \bibinfo {author} {\bibfnamefont {A.}~\bibnamefont {Shimony}}, \ and\
  \bibinfo {author} {\bibfnamefont {R.~A.}\ \bibnamefont {Holt}},\ }\href
  {\doibase 10.1103/PhysRevLett.23.880} {\bibfield  {journal} {\bibinfo
  {journal} {Phys. Rev. Lett.}\ }\textbf {\bibinfo {volume} {23}},\ \bibinfo
  {pages} {880} (\bibinfo {year} {1969})}\BibitemShut {NoStop}%
\bibitem [{\citenamefont {Le~Jeannic}\ \emph {et~al.}(2021)\citenamefont
  {Le~Jeannic}, \citenamefont {Ramos}, \citenamefont {Simonsen}, \citenamefont
  {Pregnolato}, \citenamefont {Liu}, \citenamefont {Schott}, \citenamefont
  {Wieck}, \citenamefont {Ludwig}, \citenamefont {Rotenberg}, \citenamefont
  {Garc\'{\i}a-Ripoll},\ and\ \citenamefont {Lodahl}}]{LeJeannic2021}%
  \BibitemOpen
  \bibfield  {author} {\bibinfo {author} {\bibfnamefont {H.}~\bibnamefont
  {Le~Jeannic}}, \bibinfo {author} {\bibfnamefont {T.}~\bibnamefont {Ramos}},
  \bibinfo {author} {\bibfnamefont {S.~F.}\ \bibnamefont {Simonsen}}, \bibinfo
  {author} {\bibfnamefont {T.}~\bibnamefont {Pregnolato}}, \bibinfo {author}
  {\bibfnamefont {Z.}~\bibnamefont {Liu}}, \bibinfo {author} {\bibfnamefont
  {R.}~\bibnamefont {Schott}}, \bibinfo {author} {\bibfnamefont {A.~D.}\
  \bibnamefont {Wieck}}, \bibinfo {author} {\bibfnamefont {A.}~\bibnamefont
  {Ludwig}}, \bibinfo {author} {\bibfnamefont {N.}~\bibnamefont {Rotenberg}},
  \bibinfo {author} {\bibfnamefont {J.~J.}\ \bibnamefont {Garc\'{\i}a-Ripoll}},
  \ and\ \bibinfo {author} {\bibfnamefont {P.}~\bibnamefont {Lodahl}},\ }\href
  {\doibase 10.1103/PhysRevLett.126.023603} {\bibfield  {journal} {\bibinfo
  {journal} {Phys. Rev. Lett.}\ }\textbf {\bibinfo {volume} {126}},\ \bibinfo
  {pages} {023603} (\bibinfo {year} {2021})}\BibitemShut {NoStop}%
\bibitem [{\citenamefont {Arcari}\ \emph {et~al.}(2014)\citenamefont {Arcari},
  \citenamefont {S\"ollner}, \citenamefont {Javadi}, \citenamefont
  {Lindskov~Hansen}, \citenamefont {Mahmoodian}, \citenamefont {Liu},
  \citenamefont {Thyrrestrup}, \citenamefont {Lee}, \citenamefont {Song},
  \citenamefont {Stobbe},\ and\ \citenamefont {Lodahl}}]{Arcari2014}%
  \BibitemOpen
  \bibfield  {author} {\bibinfo {author} {\bibfnamefont {M.}~\bibnamefont
  {Arcari}}, \bibinfo {author} {\bibfnamefont {I.}~\bibnamefont {S\"ollner}},
  \bibinfo {author} {\bibfnamefont {A.}~\bibnamefont {Javadi}}, \bibinfo
  {author} {\bibfnamefont {S.}~\bibnamefont {Lindskov~Hansen}}, \bibinfo
  {author} {\bibfnamefont {S.}~\bibnamefont {Mahmoodian}}, \bibinfo {author}
  {\bibfnamefont {J.}~\bibnamefont {Liu}}, \bibinfo {author} {\bibfnamefont
  {H.}~\bibnamefont {Thyrrestrup}}, \bibinfo {author} {\bibfnamefont {E.~H.}\
  \bibnamefont {Lee}}, \bibinfo {author} {\bibfnamefont {J.~D.}\ \bibnamefont
  {Song}}, \bibinfo {author} {\bibfnamefont {S.}~\bibnamefont {Stobbe}}, \ and\
  \bibinfo {author} {\bibfnamefont {P.}~\bibnamefont {Lodahl}},\ }\href
  {\doibase 10.1103/PhysRevLett.113.093603} {\bibfield  {journal} {\bibinfo
  {journal} {Phys. Rev. Lett.}\ }\textbf {\bibinfo {volume} {113}},\ \bibinfo
  {pages} {093603} (\bibinfo {year} {2014})}\BibitemShut {NoStop}%
\bibitem [{\citenamefont {Kwiat}\ \emph {et~al.}(1993)\citenamefont {Kwiat},
  \citenamefont {Steinberg},\ and\ \citenamefont {Chiao}}]{Kwiat1993}%
  \BibitemOpen
  \bibfield  {author} {\bibinfo {author} {\bibfnamefont {P.~G.}\ \bibnamefont
  {Kwiat}}, \bibinfo {author} {\bibfnamefont {A.~M.}\ \bibnamefont
  {Steinberg}}, \ and\ \bibinfo {author} {\bibfnamefont {R.~Y.}\ \bibnamefont
  {Chiao}},\ }\href {\doibase 10.1103/PhysRevA.47.R2472} {\bibfield  {journal}
  {\bibinfo  {journal} {Physical Review A}\ }\textbf {\bibinfo {volume} {47}},\
  \bibinfo {pages} {R2472} (\bibinfo {year} {1993})}\BibitemShut {NoStop}%
\bibitem [{\citenamefont {Martin}\ \emph {et~al.}(2017)\citenamefont {Martin},
  \citenamefont {Guerreiro}, \citenamefont {Tiranov}, \citenamefont
  {Designolle}, \citenamefont {Fr\"owis}, \citenamefont {Brunner},
  \citenamefont {Huber},\ and\ \citenamefont {Gisin}}]{Martin2017}%
  \BibitemOpen
  \bibfield  {author} {\bibinfo {author} {\bibfnamefont {A.}~\bibnamefont
  {Martin}}, \bibinfo {author} {\bibfnamefont {T.}~\bibnamefont {Guerreiro}},
  \bibinfo {author} {\bibfnamefont {A.}~\bibnamefont {Tiranov}}, \bibinfo
  {author} {\bibfnamefont {S.}~\bibnamefont {Designolle}}, \bibinfo {author}
  {\bibfnamefont {F.}~\bibnamefont {Fr\"owis}}, \bibinfo {author}
  {\bibfnamefont {N.}~\bibnamefont {Brunner}}, \bibinfo {author} {\bibfnamefont
  {M.}~\bibnamefont {Huber}}, \ and\ \bibinfo {author} {\bibfnamefont
  {N.}~\bibnamefont {Gisin}},\ }\href {\doibase 10.1103/PhysRevLett.118.110501}
  {\bibfield  {journal} {\bibinfo  {journal} {Phys. Rev. Lett.}\ }\textbf
  {\bibinfo {volume} {118}},\ \bibinfo {pages} {110501} (\bibinfo {year}
  {2017})}\BibitemShut {NoStop}%
\bibitem [{\citenamefont {Steinbrecher}\ \emph {et~al.}(2019)\citenamefont
  {Steinbrecher}, \citenamefont {Olson}, \citenamefont {Englund},\ and\
  \citenamefont {Carolan}}]{Steinbrecher2019}%
  \BibitemOpen
  \bibfield  {author} {\bibinfo {author} {\bibfnamefont {G.~R.}\ \bibnamefont
  {Steinbrecher}}, \bibinfo {author} {\bibfnamefont {J.~P.}\ \bibnamefont
  {Olson}}, \bibinfo {author} {\bibfnamefont {D.}~\bibnamefont {Englund}}, \
  and\ \bibinfo {author} {\bibfnamefont {J.}~\bibnamefont {Carolan}},\ }\href
  {\doibase 10.1038/s41534-019-0174-7} {\bibfield  {journal} {\bibinfo
  {journal} {npj Quantum Information}\ }\textbf {\bibinfo {volume} {5}}
  (\bibinfo {year} {2019}),\ 10.1038/s41534-019-0174-7}\BibitemShut {NoStop}%
\bibitem [{\citenamefont {Ke}\ \emph {et~al.}(2019)\citenamefont {Ke},
  \citenamefont {Poshakinskiy}, \citenamefont {Lee}, \citenamefont {Kivshar},\
  and\ \citenamefont {Poddubny}}]{Ke2019a}%
  \BibitemOpen
  \bibfield  {author} {\bibinfo {author} {\bibfnamefont {Y.}~\bibnamefont
  {Ke}}, \bibinfo {author} {\bibfnamefont {A.~V.}\ \bibnamefont
  {Poshakinskiy}}, \bibinfo {author} {\bibfnamefont {C.}~\bibnamefont {Lee}},
  \bibinfo {author} {\bibfnamefont {Y.~S.}\ \bibnamefont {Kivshar}}, \ and\
  \bibinfo {author} {\bibfnamefont {A.~N.}\ \bibnamefont {Poddubny}},\ }\href
  {\doibase 10.1103/PhysRevLett.123.253601} {\bibfield  {journal} {\bibinfo
  {journal} {Physical Review Letters}\ }\textbf {\bibinfo {volume} {123}},\
  \bibinfo {pages} {253601} (\bibinfo {year} {2019})}\BibitemShut {NoStop}%
\bibitem [{\citenamefont {Tiranov}\ \emph {et~al.}(2023)\citenamefont
  {Tiranov}, \citenamefont {Angelopoulou}, \citenamefont {{van Diepen}},
  \citenamefont {Schrinski}, \citenamefont {Sandberg}, \citenamefont {Wang},
  \citenamefont {Midolo}, \citenamefont {Scholz}, \citenamefont {Wieck},
  \citenamefont {Ludwig}, \citenamefont {S{\o}rensen},\ and\ \citenamefont
  {Lodahl}}]{Tiranov2023}%
  \BibitemOpen
  \bibfield  {author} {\bibinfo {author} {\bibfnamefont {A.}~\bibnamefont
  {Tiranov}}, \bibinfo {author} {\bibfnamefont {V.}~\bibnamefont
  {Angelopoulou}}, \bibinfo {author} {\bibfnamefont {C.~J.}\ \bibnamefont {{van
  Diepen}}}, \bibinfo {author} {\bibfnamefont {B.}~\bibnamefont {Schrinski}},
  \bibinfo {author} {\bibfnamefont {O.~A.~D.}\ \bibnamefont {Sandberg}},
  \bibinfo {author} {\bibfnamefont {Y.}~\bibnamefont {Wang}}, \bibinfo {author}
  {\bibfnamefont {L.}~\bibnamefont {Midolo}}, \bibinfo {author} {\bibfnamefont
  {S.}~\bibnamefont {Scholz}}, \bibinfo {author} {\bibfnamefont {A.~D.}\
  \bibnamefont {Wieck}}, \bibinfo {author} {\bibfnamefont {A.}~\bibnamefont
  {Ludwig}}, \bibinfo {author} {\bibfnamefont {A.~S.}\ \bibnamefont
  {S{\o}rensen}}, \ and\ \bibinfo {author} {\bibfnamefont {P.}~\bibnamefont
  {Lodahl}},\ }\href {\doibase 10.1126/science.ade9324} {\bibfield  {journal}
  {\bibinfo  {journal} {Science}\ }\textbf {\bibinfo {volume} {379}},\ \bibinfo
  {pages} {389} (\bibinfo {year} {2023})}\BibitemShut {NoStop}%
\bibitem [{\citenamefont {Alexander}\ \emph {et~al.}(2016)\citenamefont
  {Alexander}, \citenamefont {Wang}, \citenamefont {Sridhar}, \citenamefont
  {Chen}, \citenamefont {Pfister},\ and\ \citenamefont
  {Menicucci}}]{Alexander2016a}%
  \BibitemOpen
  \bibfield  {author} {\bibinfo {author} {\bibfnamefont {R.~N.}\ \bibnamefont
  {Alexander}}, \bibinfo {author} {\bibfnamefont {P.}~\bibnamefont {Wang}},
  \bibinfo {author} {\bibfnamefont {N.}~\bibnamefont {Sridhar}}, \bibinfo
  {author} {\bibfnamefont {M.}~\bibnamefont {Chen}}, \bibinfo {author}
  {\bibfnamefont {O.}~\bibnamefont {Pfister}}, \ and\ \bibinfo {author}
  {\bibfnamefont {N.~C.}\ \bibnamefont {Menicucci}},\ }\href {\doibase
  10.1103/PhysRevA.94.032327} {\bibfield  {journal} {\bibinfo  {journal} {Phys.
  Rev. A}\ }\textbf {\bibinfo {volume} {94}},\ \bibinfo {pages} {032327}
  (\bibinfo {year} {2016})}\BibitemShut {NoStop}%
\bibitem [{\citenamefont {Zhong}\ \emph {et~al.}(2015)\citenamefont {Zhong},
  \citenamefont {Zhou}, \citenamefont {Horansky}, \citenamefont {Lee},
  \citenamefont {Verma}, \citenamefont {Lita}, \citenamefont {Restelli},
  \citenamefont {Bienfang}, \citenamefont {Mirin}, \citenamefont {Gerrits},
  \citenamefont {Nam}, \citenamefont {Marsili}, \citenamefont {Shaw},
  \citenamefont {Zhang}, \citenamefont {Wang}, \citenamefont {Englund},
  \citenamefont {Wornell}, \citenamefont {Shapiro},\ and\ \citenamefont
  {Wong}}]{Zhong2015}%
  \BibitemOpen
  \bibfield  {author} {\bibinfo {author} {\bibfnamefont {T.}~\bibnamefont
  {Zhong}}, \bibinfo {author} {\bibfnamefont {H.}~\bibnamefont {Zhou}},
  \bibinfo {author} {\bibfnamefont {R.~D.}\ \bibnamefont {Horansky}}, \bibinfo
  {author} {\bibfnamefont {C.}~\bibnamefont {Lee}}, \bibinfo {author}
  {\bibfnamefont {V.~B.}\ \bibnamefont {Verma}}, \bibinfo {author}
  {\bibfnamefont {A.~E.}\ \bibnamefont {Lita}}, \bibinfo {author}
  {\bibfnamefont {A.}~\bibnamefont {Restelli}}, \bibinfo {author}
  {\bibfnamefont {J.~C.}\ \bibnamefont {Bienfang}}, \bibinfo {author}
  {\bibfnamefont {R.~P.}\ \bibnamefont {Mirin}}, \bibinfo {author}
  {\bibfnamefont {T.}~\bibnamefont {Gerrits}}, \bibinfo {author} {\bibfnamefont
  {S.~W.}\ \bibnamefont {Nam}}, \bibinfo {author} {\bibfnamefont
  {F.}~\bibnamefont {Marsili}}, \bibinfo {author} {\bibfnamefont {M.~D.}\
  \bibnamefont {Shaw}}, \bibinfo {author} {\bibfnamefont {Z.}~\bibnamefont
  {Zhang}}, \bibinfo {author} {\bibfnamefont {L.}~\bibnamefont {Wang}},
  \bibinfo {author} {\bibfnamefont {D.}~\bibnamefont {Englund}}, \bibinfo
  {author} {\bibfnamefont {G.~W.}\ \bibnamefont {Wornell}}, \bibinfo {author}
  {\bibfnamefont {J.~H.}\ \bibnamefont {Shapiro}}, \ and\ \bibinfo {author}
  {\bibfnamefont {F.~N.~C.}\ \bibnamefont {Wong}},\ }\href {\doibase
  10.1088/1367-2630/17/2/022002} {\bibfield  {journal} {\bibinfo  {journal}
  {New Journal of Physics}\ }\textbf {\bibinfo {volume} {17}},\ \bibinfo
  {pages} {022002} (\bibinfo {year} {2015})}\BibitemShut {NoStop}%
\bibitem [{\citenamefont {S\'anchez Mu\~noz}\ \emph {et~al.}(2021)\citenamefont
  {S\'anchez Mu\~noz}, \citenamefont {Frascella},\ and\ \citenamefont
  {Schlawin}}]{Carlos2021}%
  \BibitemOpen
  \bibfield  {author} {\bibinfo {author} {\bibfnamefont {C.}~\bibnamefont
  {S\'anchez Mu\~noz}}, \bibinfo {author} {\bibfnamefont {G.}~\bibnamefont
  {Frascella}}, \ and\ \bibinfo {author} {\bibfnamefont {F.}~\bibnamefont
  {Schlawin}},\ }\href {\doibase 10.1103/PhysRevResearch.3.033250} {\bibfield
  {journal} {\bibinfo  {journal} {Phys. Rev. Research}\ }\textbf {\bibinfo
  {volume} {3}},\ \bibinfo {pages} {033250} (\bibinfo {year}
  {2021})}\BibitemShut {NoStop}%
\bibitem [{\citenamefont {Kaiser}\ \emph {et~al.}(2018)\citenamefont {Kaiser},
  \citenamefont {Vergyris}, \citenamefont {Aktas}, \citenamefont {Babin},
  \citenamefont {Labont{\'e}},\ and\ \citenamefont {Tanzilli}}]{Kaiser2018}%
  \BibitemOpen
  \bibfield  {author} {\bibinfo {author} {\bibfnamefont {F.}~\bibnamefont
  {Kaiser}}, \bibinfo {author} {\bibfnamefont {P.}~\bibnamefont {Vergyris}},
  \bibinfo {author} {\bibfnamefont {D.}~\bibnamefont {Aktas}}, \bibinfo
  {author} {\bibfnamefont {C.}~\bibnamefont {Babin}}, \bibinfo {author}
  {\bibfnamefont {L.}~\bibnamefont {Labont{\'e}}}, \ and\ \bibinfo {author}
  {\bibfnamefont {S.}~\bibnamefont {Tanzilli}},\ }\href {\doibase
  10.1038/lsa.2017.163} {\bibfield  {journal} {\bibinfo  {journal} {Light:
  Science \& Applications}\ }\textbf {\bibinfo {volume} {7}},\ \bibinfo {pages}
  {17163} (\bibinfo {year} {2018})}\BibitemShut {NoStop}%
\bibitem [{\citenamefont {Zhou}\ \emph {et~al.}(2018)\citenamefont {Zhou},
  \citenamefont {Kulkova}, \citenamefont {Lund-Hansen}, \citenamefont {Hansen},
  \citenamefont {Lodahl},\ and\ \citenamefont {Midolo}}]{Zhou2018}%
  \BibitemOpen
  \bibfield  {author} {\bibinfo {author} {\bibfnamefont {X.}~\bibnamefont
  {Zhou}}, \bibinfo {author} {\bibfnamefont {I.}~\bibnamefont {Kulkova}},
  \bibinfo {author} {\bibfnamefont {T.}~\bibnamefont {Lund-Hansen}}, \bibinfo
  {author} {\bibfnamefont {S.~L.}\ \bibnamefont {Hansen}}, \bibinfo {author}
  {\bibfnamefont {P.}~\bibnamefont {Lodahl}}, \ and\ \bibinfo {author}
  {\bibfnamefont {L.}~\bibnamefont {Midolo}},\ }\href {\doibase
  10.1063/1.5055622} {\bibfield  {journal} {\bibinfo  {journal} {Applied
  Physics Letters}\ }\textbf {\bibinfo {volume} {113}},\ \bibinfo {pages}
  {251103} (\bibinfo {year} {2018})}\BibitemShut {NoStop}%
\bibitem [{\citenamefont {Uppu}\ \emph {et~al.}(2020)\citenamefont {Uppu},
  \citenamefont {Pedersen}, \citenamefont {Wang}, \citenamefont {Olesen},
  \citenamefont {Papon}, \citenamefont {Zhou}, \citenamefont {Midolo},
  \citenamefont {Scholz}, \citenamefont {Wieck}, \citenamefont {Ludwig},\ and\
  \citenamefont {Lodahl}}]{Uppu2020}%
  \BibitemOpen
  \bibfield  {author} {\bibinfo {author} {\bibfnamefont {R.}~\bibnamefont
  {Uppu}}, \bibinfo {author} {\bibfnamefont {F.~T.}\ \bibnamefont {Pedersen}},
  \bibinfo {author} {\bibfnamefont {Y.}~\bibnamefont {Wang}}, \bibinfo {author}
  {\bibfnamefont {C.~T.}\ \bibnamefont {Olesen}}, \bibinfo {author}
  {\bibfnamefont {C.}~\bibnamefont {Papon}}, \bibinfo {author} {\bibfnamefont
  {X.}~\bibnamefont {Zhou}}, \bibinfo {author} {\bibfnamefont {L.}~\bibnamefont
  {Midolo}}, \bibinfo {author} {\bibfnamefont {S.}~\bibnamefont {Scholz}},
  \bibinfo {author} {\bibfnamefont {A.~D.}\ \bibnamefont {Wieck}}, \bibinfo
  {author} {\bibfnamefont {A.}~\bibnamefont {Ludwig}}, \ and\ \bibinfo {author}
  {\bibfnamefont {P.}~\bibnamefont {Lodahl}},\ }\href {\doibase
  10.1126/sciadv.abc8268} {\bibfield  {journal} {\bibinfo  {journal} {Science
  Advances}\ }\textbf {\bibinfo {volume} {6}} (\bibinfo {year} {2020}),\
  10.1126/sciadv.abc8268}\BibitemShut {NoStop}%
\bibitem [{\citenamefont {Fan}\ \emph {et~al.}(2010)\citenamefont {Fan},
  \citenamefont {Kocaba\ifmmode~\mbox{\c{s}}\else \c{s}\fi{}},\ and\
  \citenamefont {Shen}}]{Fan2010}%
  \BibitemOpen
  \bibfield  {author} {\bibinfo {author} {\bibfnamefont {S.}~\bibnamefont
  {Fan}}, \bibinfo {author} {\bibfnamefont {i.~m. c.~E.}\ \bibnamefont
  {Kocaba\ifmmode~\mbox{\c{s}}\else \c{s}\fi{}}}, \ and\ \bibinfo {author}
  {\bibfnamefont {J.-T.}\ \bibnamefont {Shen}},\ }\href {\doibase
  10.1103/PhysRevA.82.063821} {\bibfield  {journal} {\bibinfo  {journal} {Phys.
  Rev. A}\ }\textbf {\bibinfo {volume} {82}},\ \bibinfo {pages} {063821}
  (\bibinfo {year} {2010})}\BibitemShut {NoStop}%
\bibitem [{\citenamefont {Chan}\ \emph {et~al.}(2023)\citenamefont {Chan},
  \citenamefont {Tiranov}, \citenamefont {Appel}, \citenamefont {Wang},
  \citenamefont {Midolo}, \citenamefont {Scholz}, \citenamefont {Wieck},
  \citenamefont {Ludwig}, \citenamefont {Sørensen},\ and\ \citenamefont
  {Lodahl}}]{chan_2023}%
  \BibitemOpen
  \bibfield  {author} {\bibinfo {author} {\bibfnamefont {M.~L.}\ \bibnamefont
  {Chan}}, \bibinfo {author} {\bibfnamefont {A.}~\bibnamefont {Tiranov}},
  \bibinfo {author} {\bibfnamefont {M.~H.}\ \bibnamefont {Appel}}, \bibinfo
  {author} {\bibfnamefont {Y.}~\bibnamefont {Wang}}, \bibinfo {author}
  {\bibfnamefont {L.}~\bibnamefont {Midolo}}, \bibinfo {author} {\bibfnamefont
  {S.}~\bibnamefont {Scholz}}, \bibinfo {author} {\bibfnamefont {A.~D.}\
  \bibnamefont {Wieck}}, \bibinfo {author} {\bibfnamefont {A.}~\bibnamefont
  {Ludwig}}, \bibinfo {author} {\bibfnamefont {A.~S.}\ \bibnamefont
  {Sørensen}}, \ and\ \bibinfo {author} {\bibfnamefont {P.}~\bibnamefont
  {Lodahl}},\ }\href {\doibase 10.1038/s41534-023-00717-5} {\bibfield
  {journal} {\bibinfo  {journal} {npj Quantum Information}\ }\textbf {\bibinfo
  {volume} {9}},\ \bibinfo {pages} {49} (\bibinfo {year} {2023})}\BibitemShut
  {NoStop}%
\bibitem [{\citenamefont {Manz}\ \emph {et~al.}(2007)\citenamefont {Manz},
  \citenamefont {Fernholz}, \citenamefont {Schmiedmayer},\ and\ \citenamefont
  {Pan}}]{manzCollisionalDecoherenceWriting2007}%
  \BibitemOpen
  \bibfield  {author} {\bibinfo {author} {\bibfnamefont {S.}~\bibnamefont
  {Manz}}, \bibinfo {author} {\bibfnamefont {T.}~\bibnamefont {Fernholz}},
  \bibinfo {author} {\bibfnamefont {J.}~\bibnamefont {Schmiedmayer}}, \ and\
  \bibinfo {author} {\bibfnamefont {J.-W.}\ \bibnamefont {Pan}},\ }\href
  {\doibase 10.1103/PhysRevA.75.040101} {\bibfield  {journal} {\bibinfo
  {journal} {Physical Review A}\ }\textbf {\bibinfo {volume} {75}},\ \bibinfo
  {pages} {040101} (\bibinfo {year} {2007})}\BibitemShut {NoStop}%
\bibitem [{\citenamefont {Wesenberg}\ and\ \citenamefont
  {M{\o}lmer}(2004)}]{wesenbergFieldRandomDistribution2004}%
  \BibitemOpen
  \bibfield  {author} {\bibinfo {author} {\bibfnamefont {J.~H.}\ \bibnamefont
  {Wesenberg}}\ and\ \bibinfo {author} {\bibfnamefont {K.}~\bibnamefont
  {M{\o}lmer}},\ }\href {\doibase 10.1103/PhysRevLett.93.143903} {\bibfield
  {journal} {\bibinfo  {journal} {Physical Review Letters}\ }\textbf {\bibinfo
  {volume} {93}},\ \bibinfo {pages} {143903} (\bibinfo {year}
  {2004})}\BibitemShut {NoStop}%
\end{thebibliography}%
\let\addcontentsline\oldaddcontentsline

\pagebreak
\clearpage 

\title{Supplementary Notes - \TitleName}

\maketitle
\onecolumngrid
\setcounter{equation}{0}
\setcounter{figure}{0}
\setcounter{table}{0}
\setcounter{page}{1}
\makeatletter
\renewcommand{\theequation}{S\arabic{equation}}
\renewcommand{\thefigure}{S\arabic{figure}}
\renewcommand{\bibnumfmt}[1]{[S#1]}
\renewcommand{\citenumfont}[1]{#1}
\renewcommand{\@seccntformat}[1]{%
  \csname the#1\endcsname.\quad
}

{
  \hypersetup{linkcolor=black}
  \tableofcontents
}

\setcounter{secnumdepth}{3}

\section{Spectral filtering of quantum dot emission}
\label{filter}

\begin{figure}[hbtp]
	\includegraphics[width=0.75\linewidth, trim=0.cm 0.05cm 0.cm 0.0cm,clip]{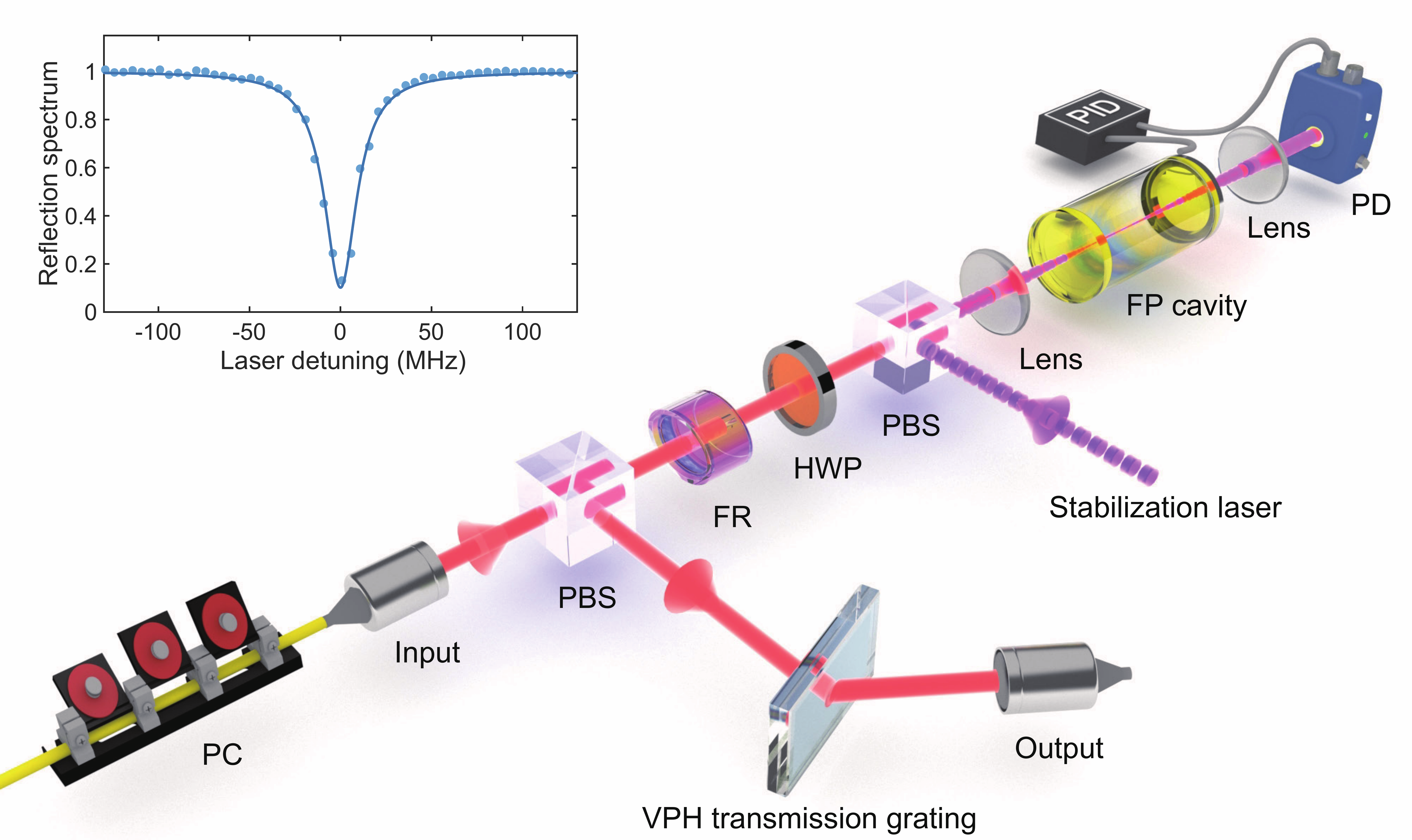}
	\caption{ 
	(color online) \textbf{Illustration of the spectral filter implemented on the QD emission.} Details of the implementation are described in the text.
	}
	\label{fig:S1}
\end{figure}

In the experiment, spectral filtering is implemented in order to purify the entanglement quality.  The filter consists of two parts: a narrow-band Fabry–Perot (FP) cavity for suppressing residual laser leakage transmitted through the QD and a volume phase holographic (VPH) transmission grating for suppressing the QD phonon sidebands, see \figref{fig:S1}. The QD-emission is prepared in the horizontal polarization with a polarization controller (PC). After passing through a Faraday rotator (FR) and a half-wave plate (HWP) at 22.5$^\circ$, the polarization remains horizontal. Subsequently, a lens is used to focus the collimated light to mode match the FP cavity. The FP cavity resonantly transmits a narrow frequency window in a linewidth of 21 MHz so that residual laser leakage (100 kHz linewidth) can be transmitted through the FP cavity. This is ensured by locking the FP cavity to the laser frequency using a PID servo-loop using a photodiode (PD) and a piezo-mounted cavity mirror. In this way, the light reflected from the FP cavity primarily consists of components inelastically scattered by the QD, while the laser component is effectively filtered. The polarization of the reflected light is rotated vertically by the magneto-optic effect in the FR and therefore reflected from the polarizing beam splitter (PBS) to obtain the output beam. After the VPH transmission grating, the first diffraction order is collected by the fiber collimator (FC).  The FP cavity is operated on resonance with the stabilization laser at low input power, and the excitation laser frequency is scanned. In this way, the reflection spectrum is recorded, see the inset of \figref{fig:S1}. By fitting the data to a Lorenzian function, we obtain an extinction of 85\% with a cavity linewidth of 21 MHz, in which the extinction is limited by minor fluctuations in the FP stabilization during the long data acquisition time, and mode matching between the input mode and intrinsic mode of the FP cavity. 

The stabilization laser is tuned 60~GHz away from the excitation laser, and the residual leakage is spectrally filtered using the VPH transmission grating with 16~GHz linewidth (\figref{fig:S1}). Both lasers are stabilized using the wavemeter up to 5~MHz precision, which limits the extinction of the FP cavity at the end.

\section{Characterization of the QD transmission intensity}
\label{RT}

\begin{figure}[hbtp]
	\includegraphics[width=0.75\linewidth, trim=0.cm 0.05cm 0.cm 0.0cm,clip]{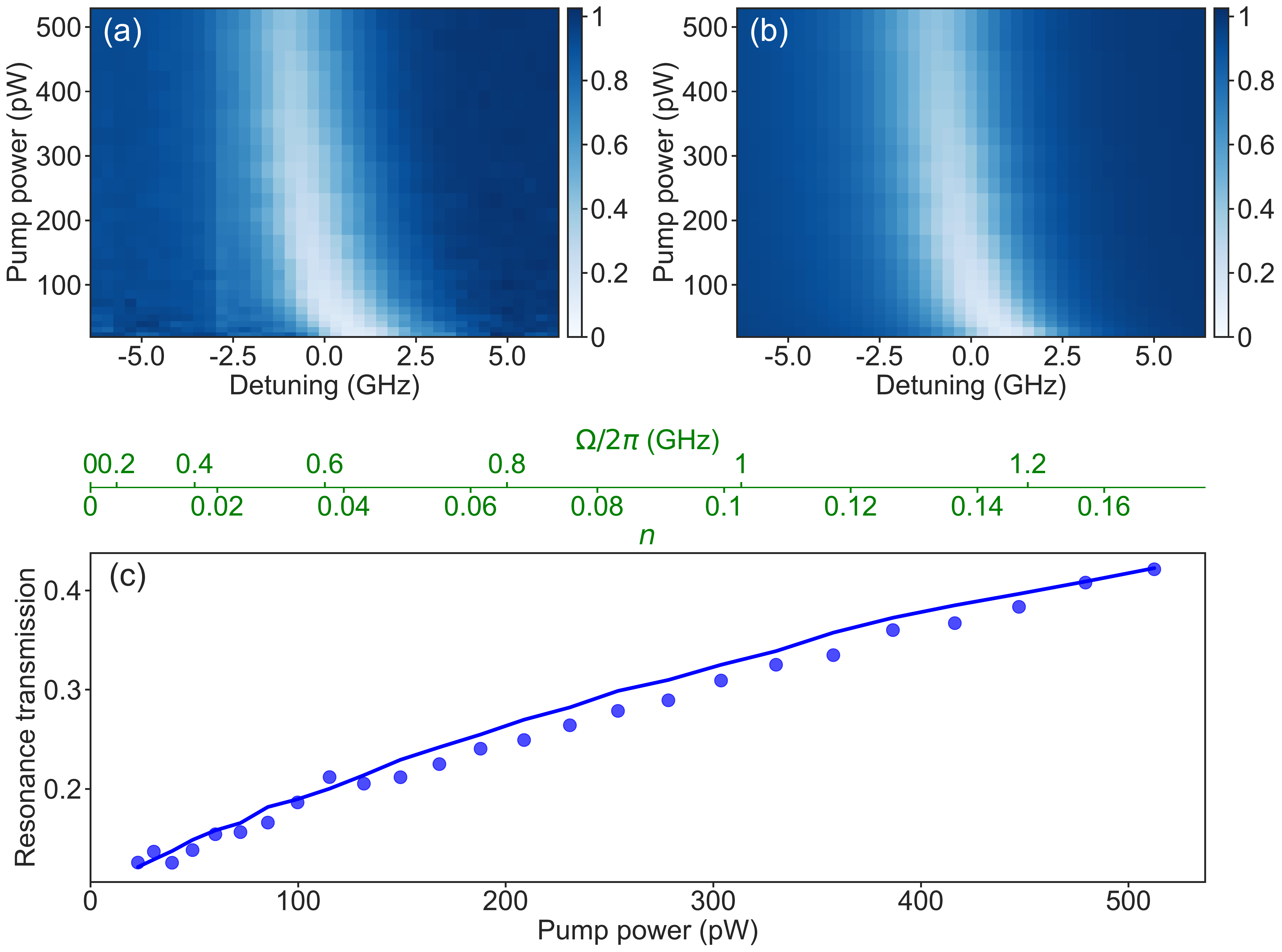}
	\caption{ 
	(color online) \textbf{The power-dependent QD transmission intensity.} (a)-(b) the measured (left) and fitted (right) transmission intensity as a function of pump power in the PhC WG and QD detuning. Based on the DC stark effect, the QD detuning relative to the laser frequency is controlled by tuning the QD bias voltage. (c). Power-dependent resonance transmission. The blue dots and curve represent the minimum transmission intensity (resonance transmission dip) extracted from (a) and (b), respectively. The two additional top x-axes represent the Rabi frequency $\Omega/2\pi$ and mean photon number $n$ within the QD lifetime, and the bottom axis shows the pump power $P_{\text{pump}}$ in the PhC WG. The extracted modelling parameters are: $\Gamma/2\pi=2.3$~GHz, $\gamma_d/2\pi=0.01$~GHz, $\sigma_{sd}/2\pi=0.39$~GHz, $\beta=0.92$, see Note \ref{Theory-QD-dynamics} for the definition of the parameters.
	}
	\label{fig:S2}
\end{figure}

The nonlinear characteristics of the QD response can be probed by recording the transmission intensity versus excitation power using one single photon detector. \figref{fig:S2} shows the measured data versus pump power and QD detuning. The transmission data are normalized to the value obtained off-resonance, i.e. for a large QD detuning. The incident laser power ($P$) is measured in the optical setup before the cryostat and subsequently converted in order to obtain the Rabi frequency $\Omega=2\sqrt{\eta P}$ or the mean photon number $n=\frac{\Omega^2}{2\beta \Gamma^2}$ in the PhC WG within a QD lifetime $\tau_{QD}$. $\eta$ is the proportionality (loss) factor connecting the measured power to the Rabi frequency. The average pump power ($P_{\text{pump}}$) in the PhC WG is estimated by $P_{\text{pump}}=nh\nu/\tau_{QD}$, where $h$ is the Planck's constant and $h\nu$ is the energy carried by a single photon at $\nu=318.6702$~THz.

\figref{fig:S2} (b) shows the theoretical fit of the experimental data from \figref{fig:S2} (a). For better visualization, the power-dependent resonance transmission is also shown in \figref{fig:S2} (c), where the blue dots and curve show the minimum transmission intensity extracted from \figref{fig:S2} (a) and (b), respectively. We extract  $\eta=0.0012 \text{ ns}^{-2}\mu\text{W}^{-1}$ and this value is used in the power-dependent measurements of the CHSH inequality (\figref{fig:2}(a)) and second-order correlations (\figref{fig:S3}). In \figref{fig:S2} (c), the conversion between the pump power $P_{\text{pump}}$ (bottom black x-axis) to $\Omega$ and $n$ (top green x-axis) is explicitly shown. For simplicity, we use only $n$ to describe the power dependence in the subsequent measurements. We find the normalized transmission intensity ($T$) measurement to be almost insensitive to the FP filter. We attribute this to the unnormalized transmitted intensity ($\text{I}$) being dominated by laser leakage, i.e., $\text{I}^{\text{Res}}_{\text{laser}}\gg\text{I}^{\text{Res}}_{\text{inelastic}}$, which is suppressed by the same extinction ratio $\eta_{\text{filter}}$ for both resonant (with superscript: ``Res") and off-resonant (``Off") data, and this mostly cancels out the effect of the filter: 

\begin{align}
    T_{\text{filtered}}=\frac{\text{I}^{\text{Res}}_{\text{inelastic}}+\text{I}^{\text{Res}}_{\text{laser}}\eta_{\text{filter}}}{\text{I}^{\text{Off}}_{\text{laser}}\eta_{\text{filter}}}\approx\frac{\text{I}^{\text{Res}}_{\text{laser}}}{\text{I}^{\text{Off}}_{\text{laser}}}\approx \frac{\text{I}^{\text{Res}}_{\text{inelastic}}+\text{I}^{\text{Res}}_{\text{laser}}}{\text{I}^{\text{Off}}_{\text{laser}}}=T_{\text{unfiltered}}.
\end{align}

\section{Two-photon correlation measurements}
\label{g2}

\begin{figure}[hbtp]
	\includegraphics[width=0.9\linewidth, trim=0.cm 0.05cm 0.cm 0.0cm,clip]{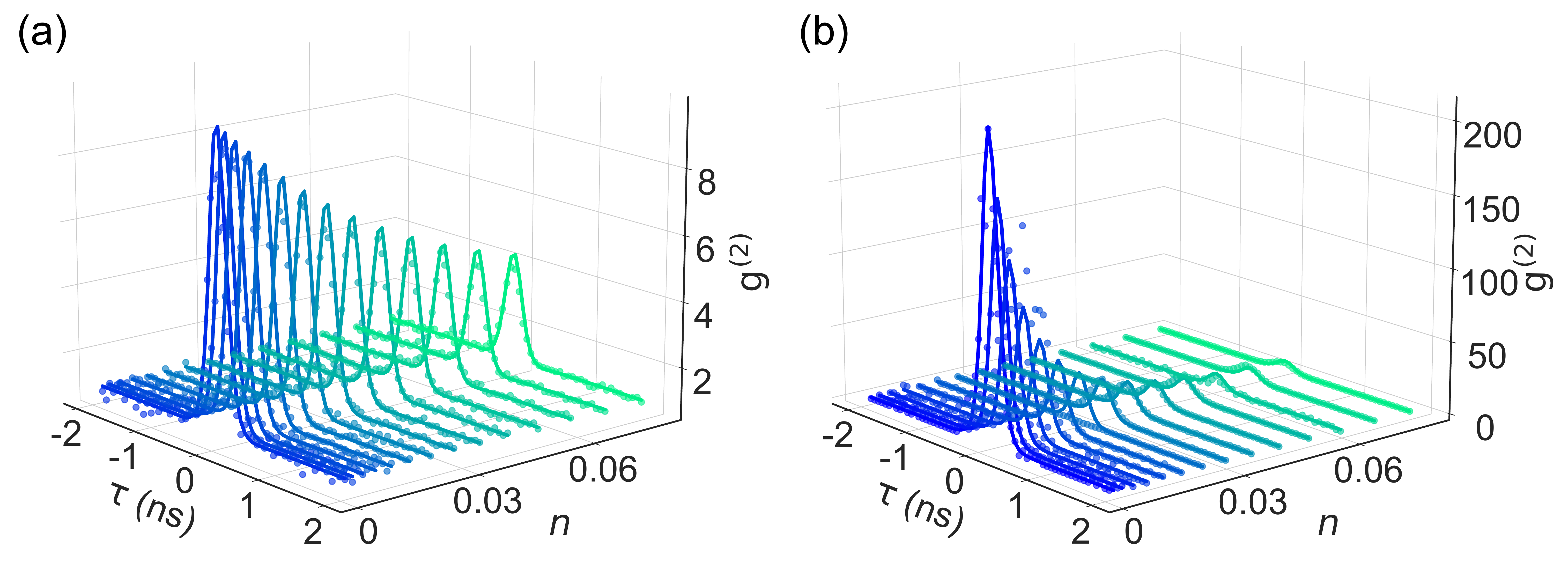}
	\caption{ 
	(color online) \textbf{Second-order correlation data as a function of time delay $\tau$ and number of incident photons $n$.} (a)-(b) $g^{(2)}(\tau)$ measurements without and with the narrow FP spectral filter. All data are normalized to the data recorded at long time delays $\tau$. 
	}
	\label{fig:S3}
\end{figure}

The two-photon correlations are measured in a Hanbury-Brown-Twiss (HBT) experiment by recording the normalized intensity correlation function $g^{(2)}(\tau)$. It is recorded using a fiber beam splitter connected directly to the QD source before the interferometer. The power dependent  $g^{(2)}(\tau)$ is recorded, see \figref{fig:S3} (a) and (b), and comparing the two different cases with and without the FP cavity filter. Pronounced bunching $(g^{(2)}(0) > 1)$ is observed in both cases, signifying the two-photon character of the light scattered by the QD. With the FP spectral notch filter, a significantly enhanced bunching is found, leading to $g^{(2)}(0)\approx210$, i.e., the filter successfully suppresses residual laser leakage which contains uncorrelated photons.  $g^{(2)}(0)$ decreases gradually with increasing pump power (quantified by $n$), which is the experimental signature of nonlinear saturation of the QD. The experimental data are fitted to the same theoretical model giving the following parameters for the unfiltered (filtered) cases: spectral diffusion of 0.52 (0.16) GHz, pure dephasing of 0.01 (0.01) GHz, coupling strength $\beta$ of 0.92 (0.96), with a time jitter of the single-photon detectors of 100 ps width and the natural linewidth of $\Gamma/2\pi=2.3$~GHz. The parameters are described in detail in section~\ref{Theory-QD-dynamics}, while the theoretical model of  $g^{(2)}(\tau)$ can be found in Ref.~\cite{LeJeannic2021}. The narrow spectral filtering improves the effective performance of the photon-emitter coupling of the QD to the photonic crystal waveguide (PhC WG).

\begin{figure}[hbtp]
	\includegraphics[width=0.55\linewidth, trim=0.cm 0.05cm 0.cm 0.0cm,clip]{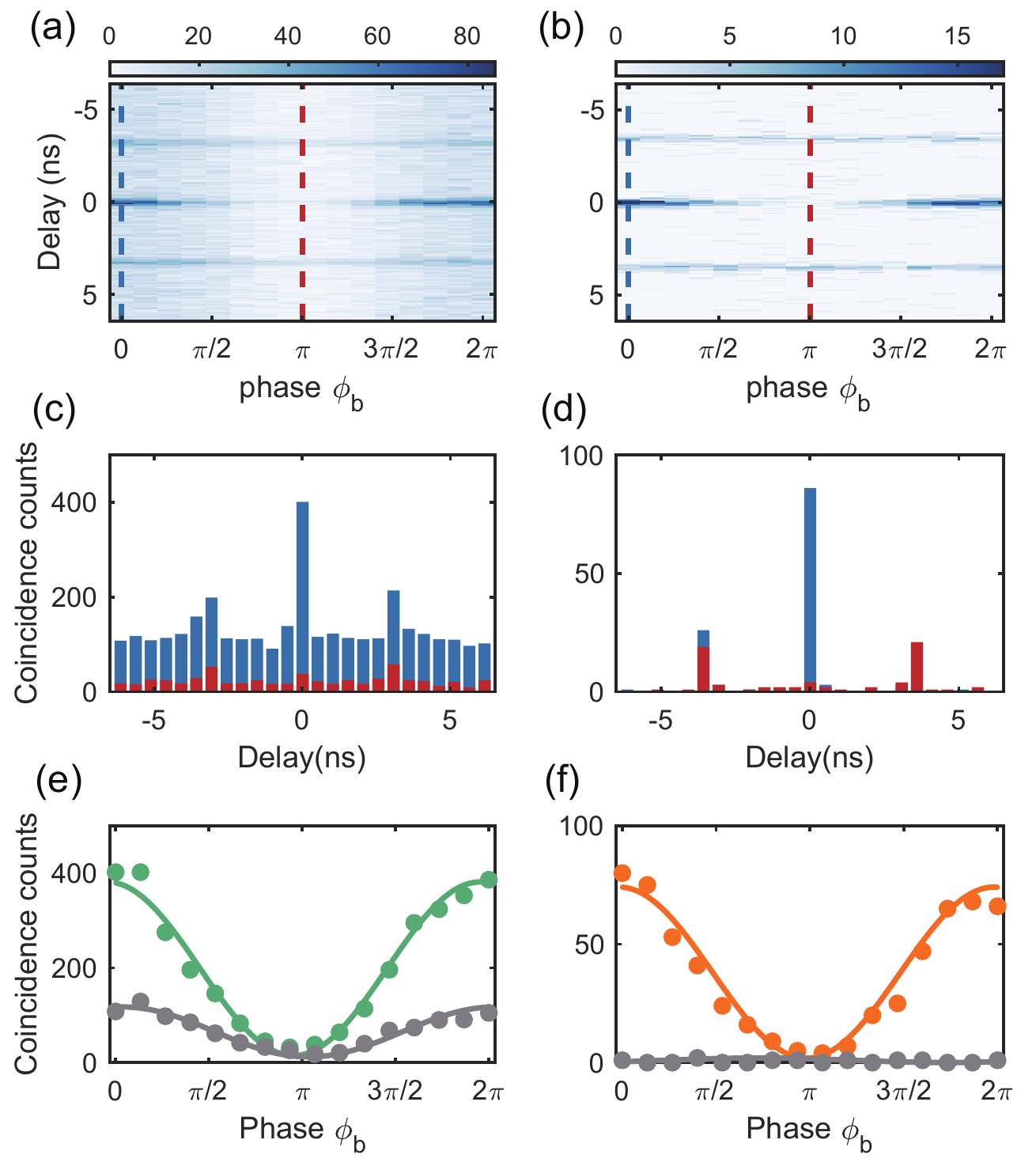}
	\caption{ 
	(color online) \textbf{Franson interference data with and without spectral filtering.} The first (second) column is the unfiltered (filtered) cases. (a)-(b) 2D coincidence interference histograms versus detector delay $\tau$ and interferometer phase $\phi_b$.  (c)-(d). Line cuts (blue and red data) extracted from (a) and (b) at the positions of the dashed line with the corresponding color. The red (blue) data correspond to destructive (constructive) interference. (e)-(f) Two-photon interference data (green and orange data) extracted from the center peak in (a) and (b) and the corresponding background (grey data) taken at $\tau = 6$ ns 
    , i.e. away from the region of the three peaks. The applied time-bin window of the coincidence data is 0.064 ns in (a)-(b) and 0.512 ns in (c)-(f). All the data from (a) to (f) is measured using the same low power, i.e., $n=0.0024$.
	}
	\label{fig:S4}
\end{figure}

\section{Additional Franson interference data}
\label{Franson}

The Franson interference data are recorded both with and without the narrow FP filter and pronounced quantum correlations are recorded in both cases, see \figref{fig:S4}. Note that here the interferometric phase $\phi_a$ is fixed to 0 during the measurement. \figref{fig:S4} (a) and (b) show the 2D coincidence interference histograms versus the time delay $\tau$ between two detectors and the interferometric phase $\phi_b$. \figref{fig:S4} (c) and (d) are the corresponding projections versus time delay comparing the two cases of constructive (blue data, $\phi_b = 0$) and destructive interference (red data, $\phi_b = \pi$). \figref{fig:S4} (e) and (f) display the interference data of the center peaks versus $\phi_b$ together with the phase-dependent background level. For the unfiltered data in the left column, the phase-dependent background is due to a small amount of laser leakage with narrow linewidth, namely the elastic photon scattering component, which can bring unwanted local classical interference when using one UMZI, and unwanted coincidence interference when using both UMZIs. Since two laser photons are always uncorrelated in time, this coincidence interference noise is time delay independent, which can be seen from the 2D interference histogram in \figref{fig:S4} (a). 

\section{Optical spectroscopy}
\label{lifetime}

\begin{figure}[hbtp]
	\includegraphics[width=0.7\linewidth, trim=0.cm 0.05cm 0.cm 0.0cm,clip]{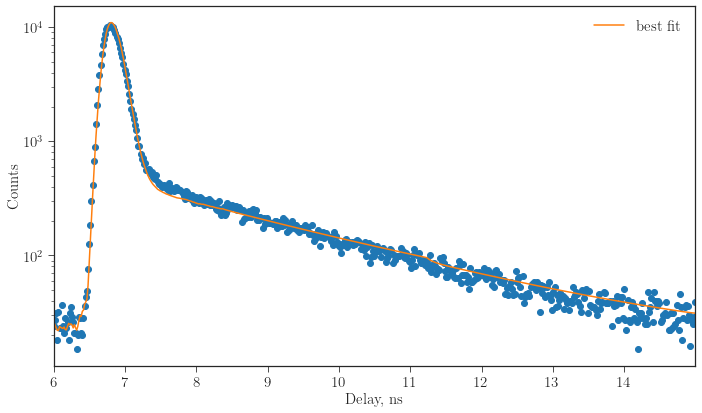}
	\caption{ 
	(color online) Lifetime measurement of the QD using p-shell excitation. Decays of two neutral excitons are visible which follow the sum of two exponentials with lifetimes of 69~ps and 2.6~ns taking into account the instrument response function.
	}
	\label{fig:Spshell}
\end{figure}

The QD lifetime is an essential parameter entering in the theoretical description of the experiment. It is extracted experimentally by exciting the QD through a higher-order transition (the p-shell) using a Ti:Sapphire mode-locked pulsed laser (pulse duration of 3~ps). The emitted fluorescence from the QD is sent into a grating filter with 25~GHZ FWHM bandwidth to remove the laser component and then detected by a single photon detector, see 
\figref{fig:Spshell}. In order to record the time-resolved emission spectra, both the detector signal and the periodic sync signal from the pulsed laser are connected to a PicoHarp time-correlator device. This allows to reconstruct the emission time of the photons giving a coincidence correlation histogram, which is determined by  the convolution of a single exponential emission decay and the instrument response function of the detector. By fitting the data, the relaxation time of the QD dipole used in this work is found to be $\tau_{QD} = \Gamma^{-1} = 69$~ps, while the lifetime of the second dipole is 2.6~ns. Since the fine structure splitting of this QD is much larger than the linewidth of the optical transition $\Gamma/2\pi$ we can effectively excite only a single dipole, such that the two-level approximation holds.

The sample is cooled down to  4~K temperature using a closed-cycle cryostat with a confocal microscope with a 0.8 numerical aperture objective. This was used to excite the QD from free space during the characterisation, and couple light into the PhC WG. Shallow etched gratings \cite{Zhou2018} allowed to couple linearly polarized light into the PhC WG within a 50 nm (FWHM) bandwidth. Details about the sample growth and fabrication can be found in \cite{Uppu2020}.

The spectroscopy of the sample containing the QD and PhC WG used in this study is described in \cite{Tiranov2023}, where it is referred as QD1.

\section{Theoretical model of two-photon scattering}
\label{scattering}
Here, we will briefly discuss the theoretical model used to describe two-photon scattering. The experiment relies on two-photon correlation measurements and the detection of a single photon in one detector conditionally prepares a quantum state within the QD lifetime $\tau_{QD}$. The energy conservation $2\omega_p = \omega_a + \omega_b$ implies an uncertainty on the energy of each scattered photon, which is proportional to $\Gamma$. This is analogous to the original Franson experiment where the decay of a three-level atom is considered~\cite{Franson1989}. We consider the resonant transmission of a weak coherent state through the 1D PhC WG. The result for the two-photon scattering can be written as \cite{Fan2010,LeJeannic2021}:

\begin{align}
    \ket{2_I} = \int d\omega^{\text{in}}_1 d\omega^{\text{in}}_2 f(\omega^{\text{in}}_1) f(\omega^{\text{in}}_2) 
    \int{d\omega_ad\omega_b} [ \chi(\omega^{\text{in}}_1) \chi(\omega^{\text{in}}_2)\delta(\omega_a - \omega^{\text{in}}_1)\delta(\omega_b - \omega^{\text{in}}_2)+ \nonumber\\
    + \frac{1}{2}\mathcal{T}_{\omega^{\text{in}}_1 \omega^{\text{in}}_2 \omega_a \omega_b} \delta(\omega_a + \omega_b - \omega^{\text{in}}_1 - \omega^{\text{in}}_2) ] \ket{1_{\omega_a}}\ket{1_{\omega_b}},
\end{align}

where $\chi(\omega)$ are the single-photon scattering corresponding to the transmission ($\chi(\omega) = t(\omega)$) and reflection  ($\chi(\omega) = r(\omega)$) coefficients and $\mathcal{T}_{\omega^{\text{in}}_1 \omega^{\text{in}}_2 \omega_a \omega_b}$ is the intrinsic two-photon correlation term.

In the case of a narrow band laser $ \omega^{\text{in}}_1 = \omega^{\text{in}}_2 = \omega_p$ and  $f(\omega) = \delta(\omega - \omega_p)$ the state in the transmission mode of the PhC WG can be written 

\begin{align}
    \ket{2_I} = \frac{1}{2} \int{d\omega_a}  \mathcal{T}_{\omega_p \omega_p \omega_a 2\omega_p-\omega_a} \ket{1_{\omega_a}}\ket{1_{2\omega_p-\omega_a}} .
\end{align}

 The term $\mathcal{T}_{\omega_p \omega_p \omega_a \omega_b}$ describes the correlated scattering, which includes the spectral entanglement. For a two-level system in a non-chiral/mirror symmetric 1D PhC WG, it can be  expressed 
 
 \begin{align}
     \mathcal{T}_{\omega_p \omega_p \omega_a 2\omega_p-\omega_a} 
     = \frac{4}{\pi\beta\Gamma} r(\omega_p)r(\omega_a)r(2\omega_p-\omega_a).
\end{align}

Using the reflection (transmission) scattering coefficients $r(\omega)= -\frac{\beta\Gamma}{\Gamma -2i(\omega-\omega_0)}$ ($t(\omega) = 1+ r(\omega)$) and resonant condition $\omega_p = \omega_0$, we can further write

 \begin{align}
     \mathcal{T}_{\omega_p \omega_p \omega_a 2\omega_p-\omega_a}=\mathcal{T}_{\Delta} & = 
      -\frac{4}{\pi\Gamma} r(\omega_a)r(2\omega_p-\omega_a) \nonumber\\
      &= -\frac{4\beta^2}{\pi\Gamma (1- 2i\Delta/\Gamma) (1+ 2i\Delta/\Gamma)} \nonumber\\
      &= -\frac{4\beta^2}{\pi\Gamma (1 + 4\frac{\Delta^2}{\Gamma^2}) }
\end{align}

 where $\Delta = \omega_a-\omega_p$ is the detuning between the generated photon and the pump laser and $\mathcal{T}_\Delta$ is the two-photon spectrum given by the Lorentzian profile. Finally, the two-photon state can be written as
 
 \begin{align}
    \ket{2_I} = \frac{1}{2} \int{d\Delta} \mathcal{T}_\Delta \ket{1_{a,\Delta}}\ket{1_{b,-\Delta}},
\end{align}

 which clearly shows the anticorrelation in energy between the two photons due to energy conservation. 
 
 The joint time distribution of two-photon fields can be described in terms of the two-photon amplitude
 
\begin{align}
     A(t,t') = \bra{0,0}E^{(+)}_a(0,t)E^{(+)}_b(0,t')\ket{2_I} = 
     \int{d\omega_a a_{\omega_a}e^{-i\omega_at}}\int{d\omega_b a_{\omega_b}e^{-i\omega_bt'}}
     \int{d\Delta\mathcal{T}_\Delta a^\dag_{\omega_p+\Delta}a^\dag_{\omega_p-\Delta}\ket{0}} \nonumber\\
     = \int{d\Delta e^{-i(\omega_p + \Delta)t}e^{-i(\omega_p - \Delta)t'}}\mathcal{T}_\Delta 
     =  e^{-i\omega_p t}e^{-i\omega_p t'}\int{d\Delta e^{i\Delta(t-t')}\mathcal{T}_\Delta}, 
\end{align}

where the last integral represents the Fourier transform of the two-photon spectral amplitude. Since each of the photons has a Lorentzian profile, the Fourier transform of their multiplication corresponds to the convolution of their individual Fourier transformations. The two-photon correlation will look as the double-exponential decay  with the decay time $\Gamma^{-1}$, defining the duration of the two-photon correlation.

\section{Quantum dot dynamics and second-order correlation functions after the interferometer}
\label{Theory-QD-dynamics}
In this section, we will describe the fields and compute the density matrix used to evaluate the second-order correlation function. We also describe how the interferometer transforms the fields such that they can be used to violate a CHSH inequality. We first write the master equation for the system, compute the steady state density matrix $\rho_{ss}$ and then apply the quantum regression theorem to the light fields and $\rho_{ss}$ to obtain the second-order correlation function $G^{(2)}(\tau)$.
\subsection{Quantum dot dynamics}
The QD is described by the Lindblad master equation for the system which reads ($\hbar = 1$) 

\begin{align}
\label{eq:rhodot}
  \dot \rho = -i \left[H , \rho \right]+ \mathcal{L}_\text{decay}[\rho] + \mathcal{L}_\text{deph}[\rho].
\end{align}

Here, the unitary part of the time evolution is given by the Hamiltonian

\begin{align}
H = \frac{\Omega}{2} \sigma_x + \Delta \sigma_{ee},
\end{align}

which accounts for the driving of the QD with Rabi frequency $\Omega$ and detuning $\Delta$. The Rabi frequency $\Omega=\Omega(t)$ may in general be time-dependent, $\sigma_x$ is the usual Pauli matrix and $\sigma_{ij}= \ket{i}\bra{j}$. 
The decay Liouvillian

\begin{align}
\mathcal{L}_\text{decay}[\rho]= \Gamma \left[ \sigma_- \rho \sigma_+ - \frac{1}{2} \left\{\sigma_+ \sigma_-, \rho \right\} \right]
\end{align}

models the total decay of the emitter $\Gamma = \gamma+ \gamma_s$ into both the waveguide mode $\gamma$ and other modes $\gamma_s$ (referred to as side modes). We additionally include a Liouvillian modelling dephasing

\begin{align}
\mathcal{L}_\text{deph}[\rho] = \frac{\gamma_d}{2} \left[ \sigma_z \rho \sigma_{z} - \rho \right],
\end{align}

where $\gamma_d$ is the dephasing rate and $\sigma_z$, $\sigma_+$ and $\sigma_-$ are the  Pauli operators. 
The ratio of the waveguide decay to the total decay defines the $\beta$-factor  $\beta = \frac{\gamma}{\Gamma}= \frac{\gamma}{\gamma+\gamma_s}$.
The transmitted field operator entering the $G^{(2)}$ calculations is $a = a_{in} \mathbb{I} - \sqrt{\beta \Gamma/2} \sigma_{ge}$, where the input field $\hat a_{in}$ is a coherent state of  amplitude $\alpha(t)$, and mean photon flux $|\alpha(t)|^2$.
The coherent field is related to the Rabi frequency by 
$\Omega/2 = \sqrt{ \beta \Gamma/2} \alpha(t)$ and thus the mean photon number within lifetime is $n = \frac{|\alpha|^2}{\Gamma} = \frac{\Omega^2}{2 \beta \Gamma^2}$.
In matrix form, the master equation reads 

\begin{align}
\dot \rho = \left(
\begin{array}{cc}
 -\Gamma  \rho _{\text{ee}}+\frac{1}{2} i \Omega  (\rho _{\text{eg}}-\rho _{\text{ge}}) & \frac{1}{2} i \Omega  (\rho _{\text{ee}}-\rho _{\text{gg}})-\frac{1}{2} \rho _{\text{eg}} (\Gamma +2 i \Delta +2 \gamma _d) \\
 -\frac{1}{2} \rho _{\text{ge}} (\Gamma -2 i \Delta +2 \gamma _d)-\frac{1}{2} i \Omega  (\rho _{\text{ee}}-\rho _{\text{gg}}) & \Gamma  \rho _{\text{ee}}-\frac{1}{2} i \Omega  (\rho _{\text{eg}}-\rho _{\text{ge}}) \\
\end{array}
\right)
\end{align}

The QD is continuously driven by a laser field, eventually reaching an equilibrium state described by the steady state density matrix $\rho_{ss}$. To find $\rho_{ss}$, we write the master equation in the form 

\begin{align}
\dot{\pmb{\rho}} = \mathbf{M} \pmb{\rho},
\end{align}

where

\begin{align}
\mathbf{M} = \left(
\begin{array}{cccc}
 -\Gamma  & \frac{i \Omega }{2} & -\frac{i \Omega}{2} & 0 \\
 \frac{i \Omega }{2} & -\frac{\Gamma }{2}-i \Delta -\gamma _d & 0 & -\frac{i \Omega}{2} \\
 -\frac{i \Omega}{2} & 0 & -\frac{\Gamma }{2}+i \Delta -\gamma _d & \frac{i \Omega }{2} \\
 \Gamma  & -\frac{i \Omega}{2} & \frac{i \Omega }{2} & 0 \\
\end{array}
\right)
\end{align}

is the coefficient matrix and

\begin{align}
\pmb{\rho} = (\rho _{\text{ee}},\rho _{\text{eg}},\rho _{\text{ge}},\rho _{\text{gg}})^{T}.
\end{align}

The master equation has a general time-dependent solution in terms of the matrix exponential

\begin{align}
\pmb{\rho}(t) =  \exp{\left( \mathbf{M} t\right)} \pmb{\rho}(0).
\end{align}

Taking $t\to \infty$ leads to an expression for $\rho_{ss}$. Alternatively, we can find the steady state (obtained when $\dot{\pmb{\rho}} = 0$) by taking the nullspace of $\mathbf{M}$ and normalizing it so that it has unit trace, which gives 

\begin{align}
\rho_{ss} = \left(
\begin{array}{cc}
 \frac{\Omega ^2 (\Gamma +2 \gamma _d)}{\Gamma  \left(4 \Delta ^2+(\Gamma +2 \gamma _d){}^2\right)+2 \Omega ^2 (\Gamma +2 \gamma _d)} & -\frac{i \Gamma  \Omega  (\Gamma -2 i \Delta +2 \gamma _d)}{\Gamma  \left(4 \Delta ^2+(\Gamma +2 \gamma _d){}^2\right)+2 \Omega ^2 (\Gamma +2 \gamma _d)} \\
 \frac{i \Gamma  \Omega  (\Gamma +2 i \Delta +2 \gamma _d)}{\Gamma  \left(4 \Delta ^2+(\Gamma +2 \gamma _d){}^2\right)+2 \Omega ^2 (\Gamma +2 \gamma _d)} & \frac{1}{\frac{\Omega ^2 (\Gamma +2 \gamma _d)}{\Gamma  \left(4 \Delta ^2+(\Gamma +2 \gamma _d){}^2\right)+\Omega ^2 (\Gamma +2 \gamma _d)}+1} \\
\end{array}
\right).
\end{align}

\subsection{Interferometer transformation}
Using an input-output formalism, we find the field that comes out of the interferometers $a$ and $b$ to be

\begin{align}
\hat a_{out}^{(a)} &= \frac{1}{\sqrt{2}} \left( a_{in}(t) + a_{in}(t-L/c) e^{-i \phi_a} \right), \\
\hat a_{out}^{(b)} &= \frac{1}{\sqrt{2}} \left( a_{in}(t) + a_{in}(t-L/c) e^{-i \phi_b} \right),
\end{align}

where a phase $\phi_{a,b}$ and a time delay $L/c$ is picked up when traveling through the long arm of the interferometer with length difference $L$ and speed of light $c$. Here
$a_{in}$ is the state that enters the interferometer, i.e., the field from the QD.
The (unnormalized) second-order correlation function after the interferometer 

\begin{align}
G^{(2)}(\tau) &=\text{Tr}\left( \hat a^{(a)}(t+\tau) \hat a^{(b)}(t) \hat \rho \hat a^{(b)}(t)^{\dag} \hat a^{(a)}(t+\tau)^{\dag} \right),
\end{align}

can be calculated using the quantum regression theorem, so that  $G^{(2)}(\tau)$ inherits its functional dependence on $\Delta, \Gamma$, etc from the density matrix, $\rho$.
We assume that $t, \tau, \frac{1}{\Gamma} \ll L/c$, so that correlations between time bins vanish, implying that $G^{(2)}$ can be factored into traces containing each separate time bin.
For the centre peak (where $t' = t+\tau$), the factored $G^{(2)}$ is 

\begin{align*}
G^{(2)}(\tau) &=2 \text{Tr}\left( a\left(t'\right)  a(t)  \rho   a(t)^{\dagger }  a\left(t'\right)^{\dagger } \right) + 2\text{Tr}\left( a \rho_{ss} a^{\dagger } \right)^2 
\\ &+ e^{-i (\phi_a+\phi_b)} \text{Tr}\left( a\left(t'\right)  a(t)  \rho\right) \text{Tr}\left( \rho   a(t)^{\dagger }  a\left(t'\right)^{\dagger}\right)
\\ &+ e^{-i(\phi_a-\phi_b)}\text{Tr}\left( a\left(t'\right)  \rho   a(t)^{\dagger } \right) \text{Tr}\left(a(t)  \rho   a\left(t'\right)^{\dagger }\right)
\\ &+ e^{-i \phi_a} \left[ \text{Tr}\left(a\rho_{ss} \right) \text{Tr}\left(a(t)  \rho   a(t)^{\dagger }  a\left(t'\right)^{\dagger } \right) + \text{Tr}\left( a\left(t'\right)  a(t)   \rho   a(t)^{\dagger }\right)\text{Tr}\left( \rho_{ss} a^{\dagger} \right) \right]
\\ &+ e^{i (\phi_a-\phi_b)} \text{Tr}\left( a(t')   \rho   a(t)^{\dagger} \right) \text{Tr}\left( a(t)    \rho   a(t')^{\dagger}\right)
\\ &+ e^{-i\phi_b} \left[ \text{Tr} \left( a(t')   \rho   a(t)^{\dagger}  a(t')^{\dagger}\right) \text{Tr}\left( a\rho_{ss} \right) + \text{Tr}\left( a(t')   a(t)   \rho   a(t)^{\dagger}\right) \text{Tr}\left( \rho_{ss} a^{\dagger}\right)  \right]  
\\ &+ e^{i(\phi_{1}+\phi_b)} \text{Tr} \left( a(t')   a(t)   \rho \right) \text{Tr} \left(  \rho   a(t)^{\dagger} a(t')^{\dagger}\right)
\\ &+ e^{i \phi_b} \left[ \text{Tr} \left( a(t') a(t)    \rho   a(t')^{\dagger}\right) \text{Tr}\left(  \rho_{ss} a^{\dagger}\right) +\text{Tr}\left( a(t')   \rho   a(t)^{\dagger} a(t')^{\dagger}\right) \text{Tr}\left( a \rho_{ss} \right) \right] 
\\ &+ e^{i\phi_a} \left[ \text{Tr} \left( a(t') a(t)    \rho   a(t)^{\dagger}\right) \text{Tr} \left(  \rho_{ss} a^{\dagger}\right) + \text{Tr} \left( a(t)    \rho   a(t)^{\dagger} a(t')^{\dagger}\right)\text{Tr}\left( a  \rho_{ss} \right)  \right] . \numberthis
\end{align*}

When $\tau = \pm L/c + \epsilon$ we obtain non-central peaks. We define $t' = t+ \epsilon$.  
The factored $G^{(2)}(\tau)$ for the non-central peaks is then

\begin{align*}
 G^{(2)}(\tau) &= \text{Tr}\left(a\left(t'\right)  a(t)  \rho   a(t)^{\dagger }  a\left(t'\right)^{\dagger }\right) + 3 \text{Tr}\left( a\rho_{ss} a^{\dagger}\right)^2
\\ &+ e^{i(\phi_a-\phi_b)} \text{Tr} \left( a \rho_{ss}\right)^{2}\text{Tr} \left(\rho   a(t)^{\dagger} a(t')^{\dagger}\right)
\\ &+ e^{i(\phi_a+\phi_b)} \text{Tr}\left(a \rho_{ss}\right) \text{Tr} \left( a(t)    \rho   a(t')^{\dagger}\right) \text{Tr}\left( \rho_{ss} a^{\dagger}\right)
\\ &+ e^{i\phi_a} \left[\text{Tr} \left( a(t')   \rho\right)\text{Tr}\left( a(t)    \rho   a(t)^{\dagger}a(t')^{\dagger}\right) + \text{Tr}\left( a \rho_{ss} \right) \text{Tr}\left( \rho_{ss} a\right) \text{Tr}\left( a(t)    \rho   a(t)^{\dagger}\right) \right] 
\\ &+ e^{-i(\phi_a+\phi_b)} \text{Tr}\left( a(t')   \rho   a(t)^{\dagger}\right) \text{Tr} \left( a(t) \rho_{ss}\right) \text{Tr} \left( \rho_{ss} a^{\dagger}\right)
\\ &+ e^{-i \phi_b}\left[ \text{Tr} \left( a(t')   \rho   a(t)^{\dagger} a(t')^{\dagger}\right) \text{Tr} \left( a(t)   \rho\right) + \text{Tr} \left( a(t')   \rho   a(t') \right) \text{Tr} \left( a \rho_{ss}\right) \text{Tr}\left(\rho_{ss} a^{\dagger}\right) \right] 
\\ &+ e^{-i(\phi_a-\phi_b)} \text{Tr} \left(a(t')   a(t)   \rho \right) \text{Tr}\left( \rho_{ss} a^{\dagger}\right)^2
\\ &+ e^{i\phi_b }\left[ \text{Tr} \left( a(t') a(t)    \rho   a(t')^{\dagger}\right) \text{Tr} \left( \rho_{ss}a^{\dagger}\right) + \text{Tr} \left( a(t')   \rho   a(t')^{\dagger}\right) \text{Tr}\left( a \rho_{ss}\right) \text{Tr}\left(\rho_{ss} a^{\dagger}\right) \right]
\\ &+ e^{-i\phi_a} \left[ \text{Tr} \left( a(t') a(t)    \rho   a(t)^{\dagger}\right) \text{Tr} \left( \rho_{ss} a^{\dagger}\right) + \text{Tr}\left( a \rho_{ss}\right) \text{Tr} \left( a(t)    \rho   a(t)^{\dagger}\right) \text{Tr}\left( \rho_{ss} a^{\dagger}\right) \right]. \numberthis
\end{align*}

\subsection{Spectral diffusion and detector time jitter}
Below, we consider two important effects relevant to modelling experimental imperfections of QD spectral diffusion and time jitter of the single-photon detector. 
Spectral diffusion corresponds to a slow (compared to the QD lifetime) frequency drift of the QD that can be modelled as a normal distribution of QD detunings:

\begin{align}
P_{SD}(\Delta,\sigma_{SD})=\frac{1}{\sqrt{2\pi \sigma_{SD}^2}} \exp\left(-\frac{\Delta^2}{2 \sigma_{SD}^2}\right) ,
\end{align}

where $\sigma_{SD}$ is the standard deviation of the spectral diffusion. We compute the effect of spectral diffusion on $G^{(2)}$ by taking the integral over all possible detunings 

\begin{align}
G^{(2)}(\sigma_{SD}, t, \Gamma, n, \beta, \gamma_d)= \int P_{SD}(\Delta,\sigma_{SD}) G^{(2)}(\Delta, t, \Gamma, n, \beta, \gamma_d) d\Delta.
\end{align}

Similarly, we consider the instrument response function (IRF) due to the time jitter of the single-photon detector, which is modelled as a Gaussian distribution:

\begin{align}
P_{IRF}(t,\sigma_{IRF})=\frac{1}{\sqrt{2\pi \sigma_{IRF}^2}} \exp\left(-\frac{t^2}{2 \sigma_{IRF}^2}\right), 
\end{align}

that influences the $G^{(2)}$ according to

\begin{align}
G^{(2)}(\sigma_{SD}, \tau, \sigma_{IRF}, \Gamma, n, \beta, \gamma_d)= \int P_{IFR}(t-\tau,\sigma_{IRF}) G^{(2)}(\sigma_{SD}, t, \Gamma, n, \beta, \gamma_d) dt
\end{align}

In the next section, the above $G^{(2)}$ expressions from the center peak will be used to construct the coincidence rates entering the CHSH inequality parameter $S$.

\subsection{CHSH Bell inequality equation as a function of $n$, $\beta$ and $\gamma_{d}$}
\label{CHSH}
In the experiment, two separate, unbalanced Mach-Zehnder interferometers (labelled $a$ and $b$) are implemented to test the CHSH inequality using the standard phase settings $\phi_a=0,\pi/2,\pi,3\pi/2$ and $\phi_b=\pi/4,3\pi/4,5\pi/4,7\pi/4$ as settings for the interferometers. We note that here two phase settings are used to mimic different output ports of each UMZI, which ideally should have separate detectors. In the experiments, however, we only use a single detector and hence we perform measurements with angles differing by $\pi$. On the other hand, the first two angles for  $\phi_a$ and $\phi_b$ could in principle be varied at will to optimize the Bell inequality violation, but for simplicity, we restrict ourselves to these angles. Experimentally, we use polarization elements (HWP and LP) to control the interferometer phase via the polarization angle (the polarization angle maps to half of the setting of the phase). Since each interferometer is applied with four different phase settings, there are 16 possible phase combinations for coincidence measurements in the central peak correlation function $G^{(2)}(\phi_a,\phi_b)$ (for brevity we here focus on the phase  and suppress the dependency of  $G^{(2)}$ on other variables). We then use the set of $G^{(2)}(\phi_a,\phi_b)$ as theoretical coincidence rates to calculate each correlation function $E(\phi_a,\phi_b)$ required in the CHSH inequality, written as:

\begin{align}
    E(\phi_a,\phi_b) = \frac{G^{(2)}(\phi_a,\phi_b)+G^{(2)}(\phi_{a\bot},\phi_{b\bot})-G^{(2)}(\phi_a,\phi_{b\bot})-G^{(2)}(\phi_{a\bot},\phi_b)}{G^{(2)}(\phi_a,\phi_b)+G^{(2)}(\phi_{a\bot},\phi_{b\bot})+G^{(2)}(\phi_a,\phi_{b\bot})+G^{(2)}(\phi_{a\bot},\phi_b)}
\end{align}

The CHSH parameter can then be determined from
\begin{align}
S= \left|E(\phi_a,\phi_b) + E(\phi_a,\phi_{b'}) - E(\phi_{a'},\phi_b) + E(\phi_{a'},\phi_{b'})\right|
\end{align}
where $\phi_{a\bot}$ and $\phi_{a'}$ represent the orthogonal (differing by $\pi$) and diagonal phase  (differing by $\pi/2$) basis  relative to $\phi_{a}$. 

To give concise analytical expressions for $S$, we will neglect spectral diffusion and detector time jitter, and assume on-resonance weak excitation of the QD  ($\Delta =0$). In this case, a complete analytical expression of $S$ can be derived but is not shown explicitly here since it is rather lengthy. For simplicity, we write the first-order Maclaurin series in $n,\gamma_d$ and in $1-\beta$

\begin{align}
    S(\Gamma,n,\beta, \gamma_d) &\approx 2 \sqrt{2}+ 32 \sqrt{2} (\beta -2) n+\gamma _d \left(\frac{224 \sqrt{2} (4-3 \beta ) n}{\Gamma }-\frac{16 \sqrt{2} (2 \beta +3)}{\Gamma }\right)
\end{align}

At the limiting values $\gamma_d =0$, $\beta=1$ and $n \to 0$, we obtain the result $2\sqrt{2}$, i.e. a maximum Bell inequality violation. We give also the exact expression for various limiting cases below. 
If we assume $\gamma_{d}=0$ and $\beta =1$, we have 

\begin{align}
S(n) =  \frac{2\sqrt2(1-4n)^2}{1+8n+32n^2} \approx 2\sqrt2\left[1-16n+O(n^2)\right]
\end{align}

where we have performed the lowest order Taylor series expansion about $n=0$.
Assuming $\gamma_{d}=0$ and $n \to 0$, we can write an exact $S$ as well as the lowest order Taylor series expansion around $\beta=1$. We obtain 

\begin{align}
S(\beta) =  \frac{2\sqrt2(1-2\beta)^2}{2-8\beta+10\beta^2-4\beta^3+\beta^4} \approx 2\sqrt2\left[1-(1-\beta)^4 +O(1-\beta)^5\right] 
\end{align}

Finally, assuming $\beta =1$, $n=0$ and performing a Taylor series expansion around $\gamma_d=0$, we obtain

\begin{align}
S(\gamma_{d}) = \frac{2 \sqrt{2} (\Gamma -2 \gamma _d){}^2}{\Gamma ^2+8 \gamma _d{}^2+4 \Gamma  \gamma _d} \approx 2\sqrt2\left[1-\frac{16 \sqrt{2}}{\Gamma}\gamma_{d}+O(\gamma_{d}^2)\right]
\end{align}

\begin{figure}[hbtp]
	\includegraphics[width=0.8\linewidth, trim=0.cm 0.05cm 0.cm 0.0cm,clip]{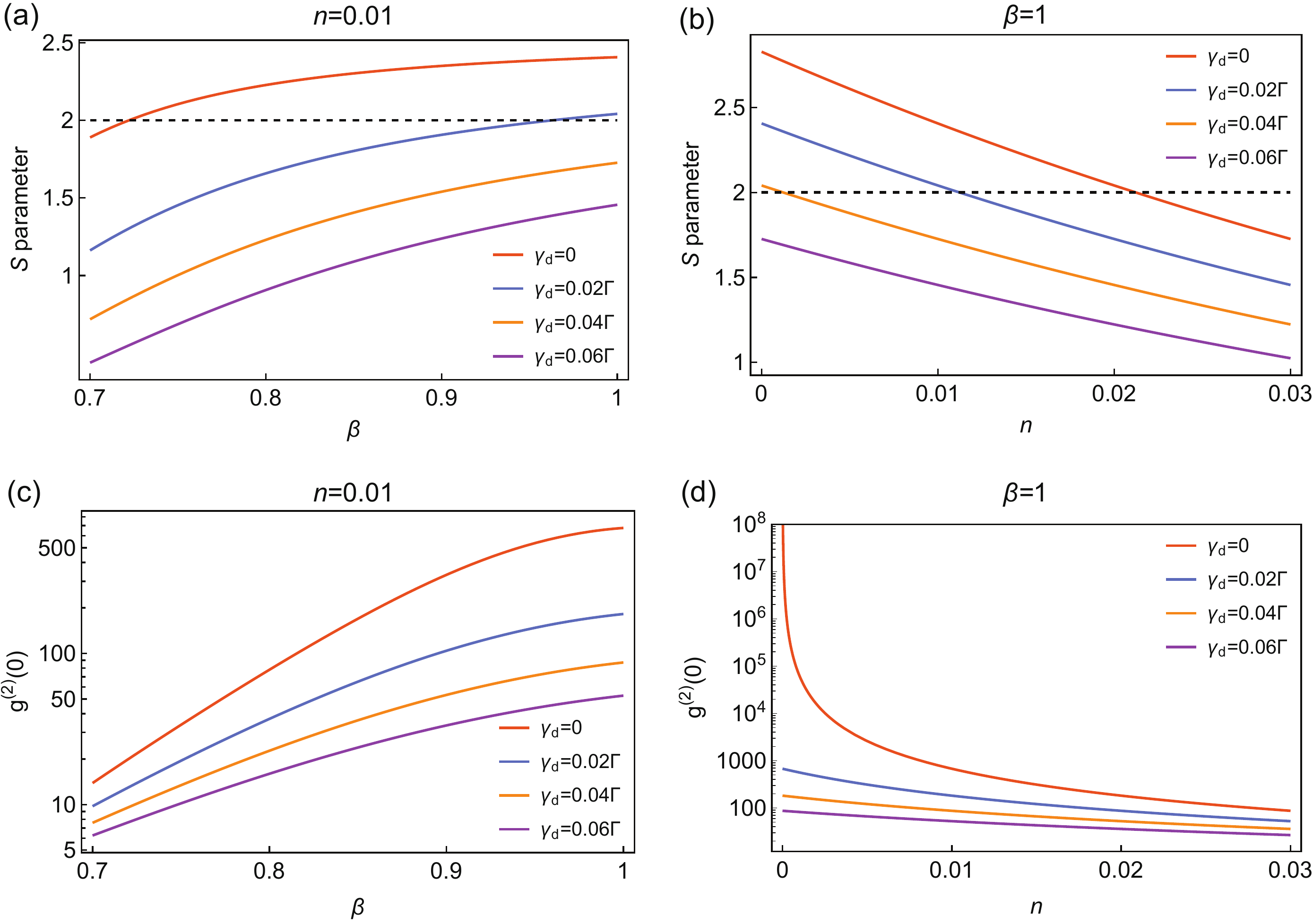}
	\caption{ 
	(color online) \textbf{$S$ parameter and $g^{(2)}(0)$ for various experimental imperfections.} (a) CHSH $S$ parameter as a function of $\beta$-factor for various values of dephasing rate $\gamma_d$ and in the low power limit ($n=0.01$). (b) CHSH $S$ parameter as a function of $n$ for various values of dephasing rate $\gamma_d$ and for $\beta=1$. (c) Second-order correlation function $g^{(2)}(0)$ as a function of $\beta$-factor for various values of the dephasing rate $\gamma_d$ and in the low power limit ($n=0.01$). (d) Second-order correlation function $g^{(2)}(0)$ versus $n$ for various values of the dephasing rate $\gamma_d$ and for $\beta=1$. The y-axis in (a)-(b) has a linear scale while the y-axis in (c)-(d) has a logarithmic scale for better visualization.
	}
	\label{fig:Sbenchmark}
\end{figure}

By evaluating the lower-order Taylor series expansions in the above equations, we can see how these parameters influence the $S$ parameter distinctly. For instance, $S(\beta)$ scales quartically with photon loss $(1-\beta)$, which makes the entanglement robust against a finite coupling efficiency.  \figref{fig:Sbenchmark} plots how $S$ depends on these imperfections based on the full theoretical model without considering spectral diffusion and time jitter.

\begin{table}[h!]
  \begin{center}
    \caption{Parameters extracted from fitting the experimental data.}
    \label{tab:table1}
    \begin{tabular}{l|S|r}
      \toprule 
      Parameter & {With FP} & Without FP \\
      \midrule 
      $\Gamma/2\pi$, GHz & 2.3 & 2.3 \\
      $\beta$ & 0.96 & 0.92 \\
      $\gamma_d/2\pi$, GHz & {$\leq$ 0.01} & $\leq$ 0.01\\
      $\sigma_{SD}/2\pi$, GHz & 0.16 & 0.39-0.52 \\
      $\sigma_{IRF}$, ps & {100} & 100 \\
      \bottomrule 
    \end{tabular}
  \end{center}
\end{table}

\subsection{Fitting the theoretical model to the data}.
\label{fit}
After the inclusion of spectral diffusion and time jitter, the observables are non-analytic functions of several parameters inherited from the density matrix and the input fields e.g., $S(\sigma_{SD},\sigma_{IRF},\Gamma,n,\beta, \gamma_d)$. To simplify the results, the delay is set to zero ($\tau=0$) in the theory, since we are interested in the coincidence rate at the center of the interference histogram. In practice, the experimental coincidence time is not infinitely narrow and the data is binned with a width of 0.512 ns for the CHSH violation. We do not consider this explicitly in the theoretical model, but we verified that artificially increasing the time jitter to take this into account did not significantly change the result. 

Our fitting procedure is conducted as follows. First, we extract $\Gamma$ by fitting data in section \ref{lifetime} and fix the obtained value for the remaining fits. An estimate of the pure dephasing rate $\gamma_d\lesssim  0.01$ GHz is obtained from a previous characterization of a similar sample \cite{chan_2023}. Based on this  we fix $\gamma_d/2\pi=0.01$ GHz for compatibility with the measurement of the Bell parameter $S$ (see below). Next, we fit the power saturation transmission data in section \ref{RT} to extract the value of the fitting parameter $\eta$, in good agreement with the power saturation $G^{(2)}$ measurement in section \ref{g2}. The remaining parameters $\sigma_{SD}$ and $\beta$ are then collectively fitted with the two aforementioned sets of power saturation data. To produce the power-dependent CHSH curve in \figref{fig:3} (c), the fitting parameters from the filtered power saturation $G^{(2)}$ data are plugged directly into our CHSH model with no additional fitting. The $S$ parameter is much more sensitive to the dephasing rate $\gamma_d$ than any of the other measurements, c.f. \figref{fig:Sbenchmark}. Any value larger than the chosen value $\gamma_d/2\pi=0.01$ GHz is incompatible with the observed value of $S$ at low $n$. Lower values of $\gamma_d$ would, however also be compatible with our measurements and the value used ($\gamma_d/2\pi=0.01$ GHz) thus provides  an upper bound.

\begin{figure}[hbtp]
	\includegraphics[width=0.8\linewidth, trim=0.cm 0.0cm 0.cm 0.0cm,clip]{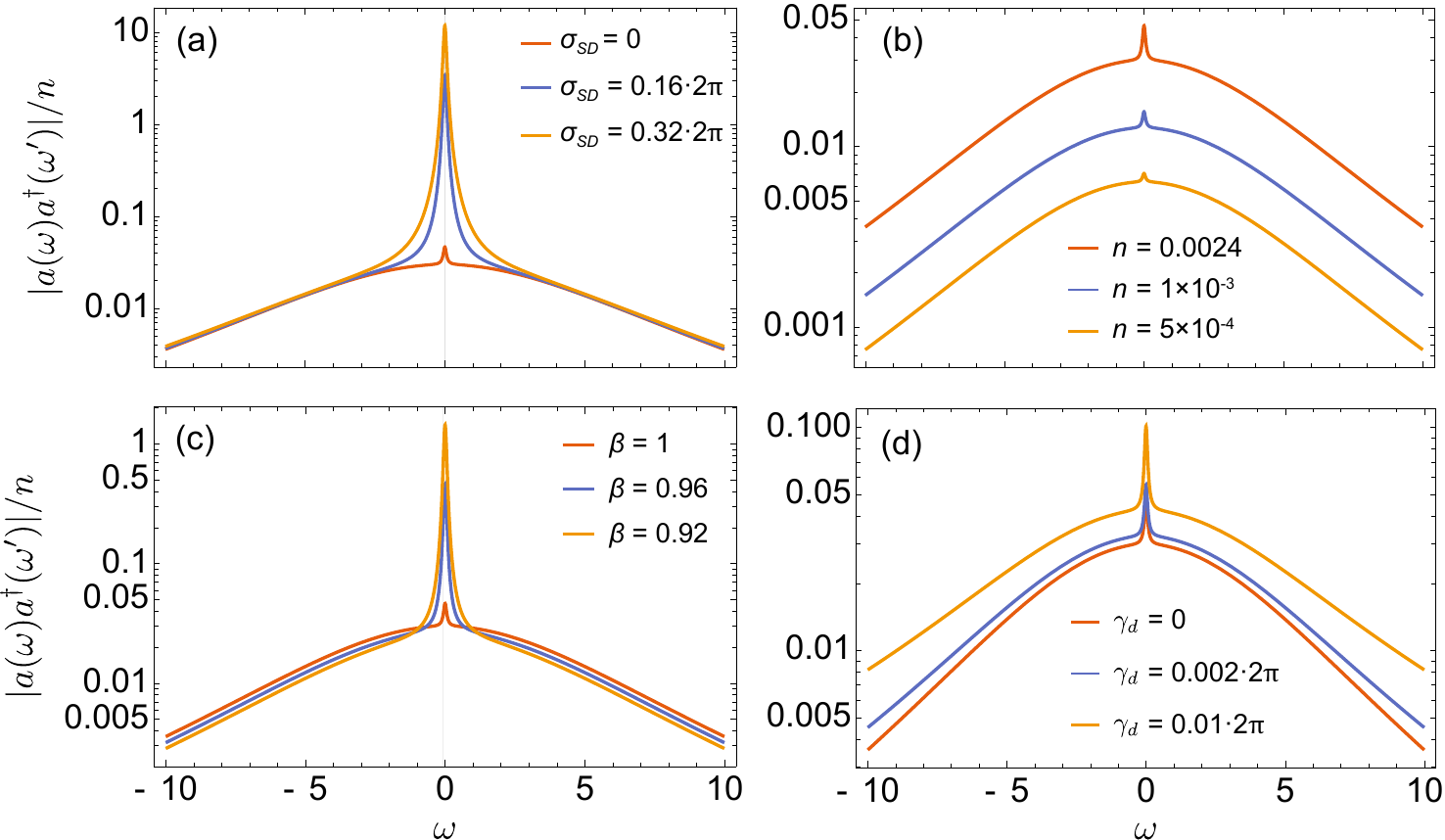}
	\caption{ 
	(color online) \textbf{Illustrative spectra $|a(\omega) a^\dag(\omega')|$ with various imperfections.} Except when otherwise stated, we use the values $n=0.0024$, $\beta=1$, $\sigma_{SD}=0$ and $\gamma_d=0$. (a) Spectra where the spectral diffusion $\sigma_{SD}$ is gradually increased. The narrow peak at $\omega=0$ is strongly affected, whilst the broad peak remains unchanged  (b) Spectra for various values of $n$. Both the central narrow peak and the broader peak are scaled up with increasing $n$.
 (c) Spectra with decreasing $\beta$. As with spectral diffusion, the dominant contribution is to increase the height of the central narrow peak whilst the broader peak is relatively unchanged.
 (d) Spectra with gradually increasing dephasing rate $\gamma_d$. The dominant effect is to increase and broaden the wide peak. The central narrow peak remains relatively unaffected.
 Note that in all plots, the width of the central peak has been scaled by a factor of 100 whilst conserving the total area so that the effect can be seen more clearly.}
\label{fig:spectra_params}
\end{figure}

We collect in Table I the parameters that are extracted from fitting the experimental data comparing with and without the FP filter. The FP filter effectively reduces leakage from the laser, which we model by varying the QD coupling parameters, in particular a modification of the effective spectral diffusion and an effective enhancement of $\beta$, whilst we keep a fixed pure dephasing rate. This can be justified by considering the illustrative spectra of the outgoing light presented in \figref{fig:spectra_params}. As seen in the figure, the outgoing spectrum consists of a narrow elastic peak and a broad inelastic emission. 
The narrow peak comes from laser leakage and vanishes in the ideal limit of vanishing imperfections ($\beta=1$, $\gamma_d=\sigma_{SD}=0$ and $n\to 0$). Under non-perfect conditions, the central narrow peak appears to quadratic order in the imperfection parameters ($1-\beta$, $\sigma_{SD}/\Gamma$, $n$ and $\gamma_d/\Gamma$) or through products of them. Since for typical parameters, $n$ and $\gamma_d/\Gamma$ ($\sim 0.01$) are much smaller than $1-\beta$ and $\sigma_{SD}/\Gamma$ ($\sim 0.1$), the latter have a much stronger influence on the coherent peak. 
In contrast to the narrow peak,   the pure dephasing induces (via the quantum jump operator $c = \sqrt{\gamma_d/2} \sigma_z$) a spectrally broad inelastic emission linear in $\gamma_d/\Gamma$ (with width $\sim \Gamma + \gamma_d$) from the QD's excited state~\cite{manzCollisionalDecoherenceWriting2007, wesenbergFieldRandomDistribution2004} and so does inelastic multi-photon scattering controlled by $n$. Upon application of the spectral filter, the narrow peak from the laser light is strongly affected, but the broad peak produced by pure dephasing and multi-photon  scattering is largely unchanged. The influence of the filter can thus be effectively captured by adjusting $\beta$ and the spectral diffusion $\sigma_{SD}$. To illustrate this, spectra where the parameters in question are changed individually are given in \figref{fig:spectra_params}.
The parameters recorded with the FP filter are obtained from fitting the data in  \figref{fig:1} (c), \figref{fig:3} (c) and \figref{fig:S3} (b). The parameters recorded without the FP filter are obtained from the transmission data in \figref{fig:1} (b), \figref{fig:S2} (a) and \figref{fig:S3} (a). It is noted that the spectral diffusion also depends on the acquisition time. Since the data in \figref{fig:1} (b) were collected with a longer integration time than the data in \figref{fig:S2} (a), a relatively larger spectral diffusion linewidth of 0.52~GHz is found compared to the other case of 0.39~GHz.

\begin{@fileswtrue}
\bibliographystyleS{apsrev4-1}
\bibliographyS{reflist_SM}
\end{@fileswtrue}

\end{document}